\documentclass[journal]{IEEEtran}
\ifCLASSINFOpdf

\else

\fi

\usepackage{amssymb}  
\usepackage{graphics} 
\usepackage{epsfig} 
\usepackage{mathptmx} 
\usepackage{times} 
\usepackage{amsmath} 
\usepackage{array} 
\usepackage{booktabs}
\usepackage{threeparttable}
\usepackage{multirow}
\usepackage{url,algorithm}
\usepackage{hyperref,algpseudocode}
\usepackage{xcolor}
\usepackage{subfigure}
\DeclareMathAlphabet{\mathcal}{OMS}{cmsy}{b}{n}
\DeclareMathAlphabet{\mathcal}{OMS}{cmsy}{m}{n}
\newcommand\numberthis{\addtocounter{equation}{1}\tag{\theequation}}
\newtheorem{theorem}{\indent Theorem}
\newtheorem{lemma}{\indent Lemma}

\newtheorem{definition}{\indent Definition}

\newtheorem{remark}{\indent Remark}
\newtheorem{problem}{\indent Problem}

\algdef{SE}[DOWHILE]{Do}{doWhile}{\algorithmicdo}[1]{\algorithmicwhile\ #1}%
\algnewcommand\algorithmicinput{\textbf{Input:}}
\algnewcommand\algorithmicoutput{\textbf{Output:}}
\algnewcommand\Input{\item[\algorithmicinput]}
\algnewcommand\Output{\item[\algorithmicoutput]}
\algrenewcommand{\algorithmiccomment}[1]{\Statex \hspace{1em}$\triangleright$ #1}

\begin{document}
\title{Generalized collective quantum tomography: algorithm design, optimization, and validation}
 \author{Shuixin Xiao,~\IEEEmembership{Member,~IEEE}, Yuanlong Wang,~\IEEEmembership{Senior Member,~IEEE},   Zhibo Hou, Aritra Das,  Ian R. Petersen,~\IEEEmembership{Life Fellow,~IEEE},  Farhad Farokhi,~\IEEEmembership{Senior Member,~IEEE}, Guo-Yong Xiang, Jie Zhao and  Daoyi Dong,~\IEEEmembership{Fellow,~IEEE}

\thanks{This research was supported by the Faculty of Engineering and Information Technology at the University of Melbourne, the Australian Research Council (DP210101938, FT220100656, DP240101494, CE170100012), the Innovation Program for Quantum Science and Technology 2023ZD0301400, and the National Natural Science Foundation of China (12288201).}
\thanks{Shuixin Xiao is with the Department of Electrical and Electronic Engineering, University of Melbourne, Parkville, VIC 3010, Australia, and School of Engineering, Australian National University, Canberra, ACT 2601, Australia (e-mail: shuixin.xiao@anu.edu.au).  }
\thanks{Yuanlong Wang is with State Key Laboratory of Mathematical Sciences, Academy of Mathematics and Systems Science, Chinese Academy of Sciences, Beijing 100190, China, and with School of Mathematical Sciences, University of Chinese Academy of Sciences, Beijing 100049, China (e-mail: wangyuanlong@amss.ac.cn).}
\thanks{Zhibo Hou and Guo-Yong Xiang are with Laboratory of Quantum Information, University of Science and Technology of China, Hefei 230026, China, and CAS Center For Excellence in Quantum Information and Quantum Physics, University of Science and Technology of China, Hefei 230026, China (e-mail: houzhibo@ustc.edu.cn, gyxiang@ustc.edu.cn). }
\thanks{Aritra Das and Jie Zhao are with
Centre for Quantum Computation and Communication Technology, Department of Quantum Science, Australian National University, Canberra, ACT 2601,
Australia (e-mail: aritra.das@anu.edu.au, jie.zhao@anu.edu.au). }
\thanks{Ian R. Petersen is with School of Engineering, Australian National University, ACT 2601, Australia (e-mail: i.r.petersen@gmail.com).}
\thanks{Farhad Farokhi is with the Department of Electrical and Electronic Engineering, University of Melbourne, Parkville, VIC 3010, Australia (e-mail: farhad.farokhi@unimelb.edu.au). }
\thanks{Daoyi Dong is with Australian Artificial Intelligence Institute, Faculty of Engineering and Information Technology, University of Technology Sydney, NSW 2007, Australia (e-mail: daoyidong@gmail.com).}
 	}

\maketitle

\begin{abstract}
Quantum tomography is a fundamental technique for characterizing, benchmarking, and verifying quantum states and devices. It plays a crucial role in advancing quantum technologies and deepening our understanding of quantum mechanics. Collective quantum state tomography, which estimates an unknown state $\rho$ through joint measurements on multiple copies $\rho\otimes\cdots\otimes\rho$ of the unknown state, offers superior information extraction efficiency. Here we extend this framework to a generalized setting where the target becomes $S_1\otimes\cdots\otimes S_n$, with each $S_i$ representing identical or distinct quantum states, detectors, or processes from the same category. We formulate these tasks as optimization problems and develop three algorithms for collective quantum state, detector and process tomography, respectively, each accompanied by an analytical characterization of the computational complexity and mean squared error (MSE) scaling.
Furthermore, we develop optimal solutions of these optimization problems using sum of squares (SOS) techniques with semi-algebraic constraints.
 The effectiveness of our proposed methods is demonstrated through numerical examples. Additionally, we experimentally demonstrate the algorithms using two-copy collective measurements, where entangled measurements directly provide information about the state purity. Compared to existing methods, our algorithms achieve lower MSEs and approach the collective MSE bound by effectively leveraging purity information.
\end{abstract}


\section{Introduction}
Quantum system identification \cite{PhysRevLett.108.080502,7130587} and quantum tomography \cite{qci,dong2022quantum} are essential for obtaining comprehensive information of quantum systems. This endeavor is critical for the thorough exploration and effective management of quantum systems \cite{dong2010quantum, wiseman2009quantum,ROUCHON2022252}. Acquiring complete models facilitates advancements in various quantum science applications, such as quantum computing \cite{qci}, quantum sensing \cite{RevModPhys.89.035002}, and quantum control \cite{Dong2023,Zhang2017}.
There are three primary tasks in quantum tomography: (i) quantum state tomography (QST), which aims to estimate unknown quantum states \cite{qstmle,Qi2013,effqst,6636034,6942238}; (ii) quantum detector tomography (QDT), which focuses on identifying and calibrating quantum measurement devices \cite{Grandi_2017,PhysRevA.64.024102,Lundeen2009,wang2019twostage}; and (iii) quantum process tomography (QPT), designed to determine the parameters characterizing unknown quantum processes \cite{qptre,surawy2021projected,xiao2024,PhysRevA.82.042307,8022944,10089832}. State, process, and detector tomography, closely connected in both mathematical and physical senses, together form an essential triad necessary to fully characterize typical quantum measurement experiments \cite{Lundeen2009}.

For QST, various algorithms have beem employed, such as Maximum Likelihood Estimation  \cite{qstmle,effqst} and Linear Regression Estimation \cite{Qi2013,xiaorank}. For low-rank quantum states, innovative approaches such as the application of compressed sensing in QST have been proposed \cite{Gross2010,Flammia2012}. Additionally, regularization techniques to enhance the accuracy of QST were introduced in \cite{MU2020108837}. To predict properties of quantum states with only a few measurements, shadow tomography methods have been proposed in \cite{Aaronson2018,Huang2020}.
For QDT, the pioneering solution method was Maximum Likelihood Estimation \cite{PhysRevA.64.024102}. Subsequent methodologies include linear regression \cite{Grandi_2017} and convex optimization \cite{Lundeen2009}. A recent advance in this field is a two-stage estimation method characterized by analytical computational complexity and an upper bound for the mean squared error (MSE) \cite{wang2019twostage}. Building upon this two-stage approach, Ref. \cite{xiao2021optimal} introduced an optimization of the probe states, while Ref. \cite{Xiao2023} examined regularization techniques in QDT.
For QPT, three main approaches are typically used, depending on the system architecture: Standard Quantum Process Tomography \cite{qci,PhysRevA.63.020101,xiaocdc2}, Ancilla-Assisted Process Tomography \cite{aaqpt,xiaoaapt}, and Direct Characterization of Quantum Dynamics \cite{dcqd}.

One appealing approach to quantum tomography is collective tomography, although its applications have so far been largely restricted to QST. Collective measurements on multiple identically prepared state copies can extract more information than individual measurements (each performed on a single copy of the state), thereby enhancing information-processing efficiency \cite{PhysRevLett.120.030404,7956181}.
Refs.~\cite{PhysRevLett.97.130501,zhu} demonstrated an enhancement in fidelity using collective measurements over separable methods in estimating mixed qubit states, and Ref.~\cite{das2024holevo} showed that, in quantum state 
tomography, collective measurements can achieve up to a $d$-fold improvement 
over separable ones, where $d$ is the system dimension. 
Furthermore, collective symmetric informationally complete measurements were 
investigated in~\cite{PhysRevLett.120.030404}, and experimentally demonstrated in optics using two-copy collective 
measurements in~\cite{Hou2018,11158864}. 
Ref.~\cite{PhysRevA.111.062409} proposed genuine collective measurements for QST, which was realized in an optical system through a three-copy experiment~\cite{zhou2023experimental}. 
More recently, Ref.~\cite{conlon2023} demonstrated provably optimal individual and two-copy collective measurements on superconducting, trapped-ion, and photonic platforms.
 Additionally, Ref. \cite{yuan2020} explored a collective measurement scheme for estimating the amount of coherence in quantum states.
Moreover, collective tomography can help reduce the back action of quantum measurement \cite{PhysRevLett.118.070601}, which has been demonstrated in optical systems~\cite{wu2019}.

Despite the above progress, existing research on collective QST has largely been restricted to scenarios where multiple copies of the quantum states are identical as $\rho\otimes\cdots\otimes\rho$, while collective tomography approaches to QDT and QPT remain relatively unexplored. To address these gaps, we propose a generalized collective tomography uniting QST, QDT, and QPT into a general framework considering
$S_1\otimes\cdots\otimes S_n$, with each $S_i$ representing identical or distinct states, detectors, or processes from the same category.
 When $S_i\neq S_j$ holds for $i\neq j$, we refer to the task as D-QST, D-QDT, or D-QPT where D implies \emph{distinct}, and when $S_1=\cdots=S_n$, we refer to it as I-QST, I-QDT, or I-QPT where I stands for \emph{identical}.  

We formulate these tasks as optimization problems with a natural tensor structure for the unknown parameters, which exhibit non-convexity and bilinearity. Building on this structure, we derive closed-form solutions tailored to each case, and analyze both their computational complexities and MSE scalings, which generalizes the previous result on QST~\cite{11158864} to all three classes of quantum tomography tasks. The complexities are shown to be comparable to those of conventional algorithms for individual QST, QDT, and QPT.

A limitation of the closed-form method is that it may yield suboptimal 
solutions. To address this issue, we reformulate the problems as 
sum-of-squares (SOS) optimization problems with semi-algebraic constraints and 
solve them using SOSTOOLS~\cite{sostools}.
 This framework provides a rigorous lower bound on the cost function. If the optimizer achieves the lower bound, the bound corresponds to the global minimum. Otherwise, only the bound is returned. To the best of our knowledge, this work represents the first application of SOS optimization techniques in collective tomography, marking a significant methodological advance.  

As benchmark cases, we apply our tomography methods to estimate
 special cases including pure states, projective measurements, and unitary processes. Numerical experiments confirm the effectiveness of the proposed methods: the closed-form approach is computationally efficient but less accurate, whereas SOS optimization achieves significantly higher accuracy at the expense of increased computational demand.  

In addition, we validate our algorithms using experimental data from~\cite{Hou2018}, which demonstrated two-copy collective measurements in QST via photonic quantum walks. The entangled measurements in this setup yield information about the purity of the quantum state, consistent with earlier theoretical~\cite{poly} and experimental~\cite{PhysRevA.75.012104} results for qubit systems. In contrast, individual measurements cannot directly capture nonlinear properties such as purity. Compared to the modified accelerated projected-gradient algorithm of~\cite{Hou2018}, our methods achieve lower MSEs and approach the theoretical collective MSE bound by exploiting this purity information.  

The main contributions of this paper are summarized as follows:  
\begin{enumerate}
    \item[(i)] We extend collective QST to a generalized collective tomography framework that unifies QST, QDT, and QPT, formulating all of these tasks as tensor-structured optimization problems.  
    \item[(ii)] We derive closed-form solutions, analyze their computational complexities and MSE scalings, and also propose SOS formulations that yield optimal solutions under semi-algebraic constraints. 
    \item[(iii)] We provide illustrative examples, including pure states, projective measurements, and unitary processes, and verify the theoretical error analysis through numerical simulations.  
    \item[(iv)] We validate the proposed methods using experimental two-copy collective QST data from~\cite{Hou2018}, demonstrating that our algorithms achieve lower MSEs than existing approaches and approach the collective MSE bound by exploiting purity information.  
\end{enumerate}

The remainder of this paper is organized as follows.  
Section~\ref{sec2} formulates the problem of implementing generalized collective QST, QDT, and QPT.  
Section~\ref{closed} presents the closed-form algorithms and presents their computational complexities and MSE scalings.  
Section~\ref{sosop} introduces the SOS optimization framework for collective QST, QDT and QPT, and Section~\ref{example} provides both conceptual demonstrations and numerical simulations.  
Section~\ref{sec7} validates the proposed methods using experimental data.  
Finally, Section~\ref{sec8} concludes the paper.

Notation:  The $ i $-th row and $ j $-th column of a matrix $ X $ is $(X)_{ij} $. The $ j $-th column of $ X $ is $ \operatorname{col}_{j}(X) $.  The transpose of $X$ is $X^T$. The conjugate $ (*) $ and transpose of $X$ is $X^\dagger$. The rank of a matrix $X$ is $\operatorname{rank}(X)$. The elements from the $m$-th to the $n$-th position ($m\leq n$) in a vector $x$ are denoted as $x_{m:n}$. The sets of  real and complex numbers are $\mathbb{R}$ and $\mathbb{C}$, respectively. The sets of  $d$-dimension real/complex vectors and  $d\times d$ real/complex matrices are $\mathbb{R}^d/\mathbb{C}^d$ and $\mathbb{R}^{d\times d}/\mathbb{C}^{d\times d}$, respectively. The identity matrix is $ I_d $ in the $d$ dimension. The imaginary unit is denoted by $\rm i=\sqrt{-1}$. 
The trace of $X$ is $\text{Tr}(X)$. The Frobenius norm of a matrix $X$ is denoted as $||X||$ and the 2-norm of a vector $ x $ is $||x||$.   The estimate of $X$ is $ \hat X $. The inner product of two matrices $X$ and $Y$ is defined as $\langle X, Y\rangle\triangleq\text{Tr}(X^\dagger Y)$.
The inner product of two vectors $x$ and $y$ is defined as $\langle x,
y\rangle\triangleq x^\dagger y$. The tensor product of $A$ and $B$ is denoted $A\otimes B$.  $A^{\otimes n}$ denotes the tensor product of $A$ with itself $n$ times. Hilbert space is $\mathbb{H}$. $\operatorname{Tr}_1(X)$ denotes the partial trace on the space $\mathbb H_1$ with $X$ belonging to the space $\mathbb H_1\otimes \mathbb H_2$. The Kronecker delta function is $\delta$. $\operatorname{diag}(a)$ for a vector $a$ denotes a diagonal matrix with the $i$-th diagonal element being the $i$-th element of the vector $a$, and  $\operatorname{diag}(A)$ represents a diagonal matrix whose diagonal elements are the same as those of the square matrix $A$.
The Pauli matrices are
 \begin{equation*}
\sigma_x=\begin{bmatrix}
0 & 1 \\
1 & 0
\end{bmatrix},\;\;
\sigma_y=\begin{bmatrix}
0 & -\mathrm{i} \\
\mathrm{i} & 0
\end{bmatrix},\;\;
\sigma_z=\begin{bmatrix}
1 & 0 \\
0 & -1
\end{bmatrix}.
\end{equation*}

\begin{table}[htbp]
	\centering
	\caption{Key symbols.}
	\label{tablet1}
	\renewcommand{\arraystretch}{1.3}
	\begin{tabular}{ll}
		\hline
		\textbf{Symbol} & \textbf{Description} \\
		\hline
		$N$ & Total number of state copies used in QST/QDT/QPT \\
		$\rho$ & True quantum state (density operator) \\
		$\tilde{\rho}, \bar{\rho}, \hat{\rho} $ & Intermediate and final estimates of $\rho$ \\
		$\mathcal{S}_{d}$ & Set of physical quantum states in dimension $d$ \\
		$P_i$ & $i$-th true POVM element \\
		$\tilde{P}_i,\bar{P}_i,\hat{P}_i $ & Intermediate and final estimates of $P_i$ \\
		$\mathcal{D}_{d}$ & Set of physical detectors in dimension $d$ \\
		$X$ &  True process matrix of the quantum process $\mathcal{E}$ \\
		$\tilde{X}, \bar{X}, \hat{X} $ & Intermediate and final estimates of $X$ \\
		$\mathcal{P}_{d}^{\operatorname{TP}}$ & Set of physical trace-preserving quantum processes \\
		$\mathcal{P}_{d}^{\neg\operatorname{TP}}$ & Set of physical non-trace-preserving quantum processes \\
		$\mathcal{R}(A)$ & Permuted version of a matrix $A$ \cite{LOAN200085} \\
		\hline
	\end{tabular}
\end{table}

\section{Problem formulation}\label{sec2}
\subsection{Preliminary knowledge}
For a matrix $A_{m\times n}$, we introduce the vectorization function:
\begin{equation}
	\begin{aligned}
		\operatorname{vec}(A_{m\times n})\triangleq&[(A)_{11},(A)_{21},\cdots,(A)_{m1},(A)_{12},\cdots,(A)_{m2},\\
		&\cdots,(A)_{1n},\cdots,(A)_{mn}]^T.
	\end{aligned}
\end{equation}
Similarly, $\operatorname{vec}^{-1}(\cdot)$ maps a $d^2\times 1$ vector into a $d\times d$ square matrix.
One common property of $\operatorname{vec}(\cdot) $ is \cite{watrous2018theory}:
\begin{equation}\label{property2}
	\operatorname{vec}(ABC)=(C^T\otimes A)\operatorname{vec}(B).
\end{equation}

A quantum state $ \rho \in \mathbb{C}^{d\times d} $  must satisfy $ \rho=\rho^{\dagger} $, $ \rho\geq0 $ and $ \operatorname{Tr}(\rho)=1 $. When $\rho=|\psi\rangle\langle\psi|$ where $|\psi\rangle\in\mathbb{C}^{d}$ is a unit vector, $ \rho $ is called a pure state. Otherwise, $\rho$ is a mixed state. Here, we define the following set to characterize all physical quantum states in a $d$-dimensional Hilbert space:
\begin{equation}
	\mathcal{S}_{d}=\left\{\rho \in \mathbb{C}^{d\times d}:  \rho=\rho^{\dagger}, \rho\geq 0,\operatorname{Tr}(\rho)=1 \right\}.
\end{equation}

In quantum physics, measurement is ubiquitous, and the measurement device is called a detector, which can be characterized by a set of measurement operators denoted as $\{P_l\}_{l=1}^{L}$. These $L$ operators form a Positive-Operator-Valued Measure (POVM), where each POVM element $P_l \in \mathbb{C}^{d \times d}$ adheres to the conditions $ P_l = P_l^{\dagger} $ and $ P_l \geq 0 $. Furthermore, they satisfy the completeness constraint $\sum_{l=1}^L P_l = I_{d}$.  Here, we define the following set to characterize all physical 
	$d$-dimensional detectors:
\begin{equation}
	\mathcal{D}_{d}=\Big\{\{P_l\}_{l}: P_l \in \mathbb{C}^{d\times d}, P_l=P_l^{\dagger}, P_l\geq 0, \sum_{l} P_l = I_{d} \Big\}.
\end{equation}

When a measurement operator $P_l$ is applied to a quantum state $\rho$, the probability of obtaining the corresponding result is governed by Born's rule \cite{qci}:
\begin{equation}
	p_{l} = \operatorname{Tr}\left(P_{l} \rho\right).
\end{equation}
From the completeness constraint, we have $\sum_{l=1}^{L} p_l=1$. In practical experiments, suppose that $N$ identical copies of $\rho$ are consumed, and the $l$-th operator occurs $N_l$ times. Then $\hat p_l=N_l/N$ serves as the experimental estimate of the true value $p_l$, with the associated measurement error denoted by $ e_{l} = \hat p_{l} - p_{l} $ \cite{wang2019twostage}. According to the central limit theorem, the distribution of $ e_{l} $ converges to a normal distribution with mean zero and variance $(p_{l} - p_{l}^{2})/N$ \cite{Qi2013, MU2020108837}. Based on the measurement data $\{p_l\}$, the target of QST (QDT) is to identify the unknown state (detector) using known POVM (measured states, also called probe/input states).

For a $ d $-dimensional quantum system, its dynamics can be described by a completely-positive (CP) linear map
$ \mathcal{E} $ and QPT aims to identify the unknown $ \mathcal{E} $.
If we input a quantum state $\rho^{ \text{in}}\in \mathbb{C}^{d \times d}$, using
Kraus operator-sum representation \cite{qci}, the output state $\rho^{ \text{out}}$ is given by 
\begin{equation}\label{Kraus}
	\rho^{ \text{out }}=\mathcal{E}(\rho^{ \text{in }})=\sum_{i=1}^{d^2} A_{i} \rho^{\text{in }}  A_{i}^{\dagger},
\end{equation}
where  $ A_{i} \in \mathbb{C}^{d \times d}$ and they satisfy
\begin{equation}\label{aleq}
	\sum_{i=1}^{d^2}  A_{i}^{\dagger} A_{i}\leq I_d.
\end{equation}

We also call $d$ the dimension of $\mathcal{E}$.
Choosing $\{E_i\}_{i=1}^{d^2}$  as the natural basis  $\{|j\rangle\langle k|\}_{1\leq j,k\leq d}$ \cite{qci,8022944} where $i=(j-1)d+k $ and $\{|j\rangle\}_{j=1}^{d}$ represents the standard basis in $\mathbb{C}^{d}$. Consequently, the natural basis $\{|j\rangle\langle k|\}_{1\leq j,k\leq d}$ is also a standard basis in $\mathbb{C}^{d\times d}$. We expand  $\left\{ A_i\right\}_{i=1}^{d^2}$ as
\begin{equation}
	{A}_i=\sum_{j=1}^{d^2} c_{{ij}} E_j.
\end{equation}
Define matrix $ (C)_{ij}\triangleq c_{{ij}} $, and in fact we have
\begin{equation}\label{cc}
	C=\left[\operatorname{vec}(A_1^{T}), \cdots, \operatorname{vec}(A_{d^2}^{T})\right]^T.
\end{equation}
Also define the matrix $ X $ as $ X\triangleq C^T C^* $, which is called process matrix \cite{qci,8022944} and $ X \in \mathbb{C}^{d^2\times d^2}$  is in one-to-one correspondence with $ \mathcal{E} $. In addition, it satisfies $X=X^{\dagger}, X\geq0, \operatorname{Tr}_{1}\left(X\right)\leq I_d $. When the equality in \eqref{aleq} holds, we have $ \operatorname{Tr}_{1}\left(X\right)= I_d $ \cite{qci,8022944} and the process $ \mathcal{E}  $ or $ X $ is trace-preserving (TP). Otherwise, the process is non-trace-preserving (non-TP) which has indeed been demonstrated in experiment \cite{PhysRevA.82.042307}. We further define two sets to characterize all TP and non-TP  process matrices of  dimension $d$:
\begin{equation}
	\mathcal{P}_{d}^{\operatorname{TP}}=\left\{X \in \mathbb{C}^{d^{2}\times d^{2}}:  X=X^{\dagger}, X\geq 0,\operatorname{Tr}_{1}(X)= I_{d} \right\},
\end{equation}
and
 \begin{equation}
 	\mathcal{P}_{d}^{\neg\operatorname{TP}}=\left\{X \in \mathbb{C}^{d^{2}\times d^{2}}:  X=X^{\dagger}, X\geq 0,\operatorname{Tr}_{1}(X)\leq  I_{d} \right\}.
 \end{equation}
The target for QPT is to identify the unknown process matrix $X $ using the known input states $ \{\rho_m^{\operatorname{in}}\}_{m=1}^{M } $ and the measurement operators $ \{P_l\}_{l=1}^{L} $, where $M$ and $ L $ are the  numbers of different kinds of  input states and measurement operators, respectively.

State, process, and detector tomography constitute a complete triad needed to fully define an experiment.
Since $\rho$, $\{P_l\}_{l=1}^{L}$ and $X$ are all Hermitian matrices, we introduce a complete basis set of orthonormal operators $\{\Omega_{j}\}_{j=1}^{d^{2}}$ in the dimension $d$, satisfying $\operatorname{Tr}\left(\Omega_{i}^{\dagger} \Omega_{j}\right) = \delta_{i j}$. Each operator $\Omega_{j}$ is Hermitian, and $\operatorname{Tr}\left(\Omega_{j}\right) = 0$ for all $j$ except $\Omega_{1} = I / \sqrt{d}$. This basis is employed in QST and QDT. For QPT, we utilize the natural basis  $\{|j\rangle\langle k|\}_{1\leq j,k\leq d}$.

In the following subsections, we propose generalized collective tomography for quantum states, detectors and quantum processes.

\subsection{Generalized collective QST}

\begin{figure}
	\centering
	\includegraphics[width=3.4in]{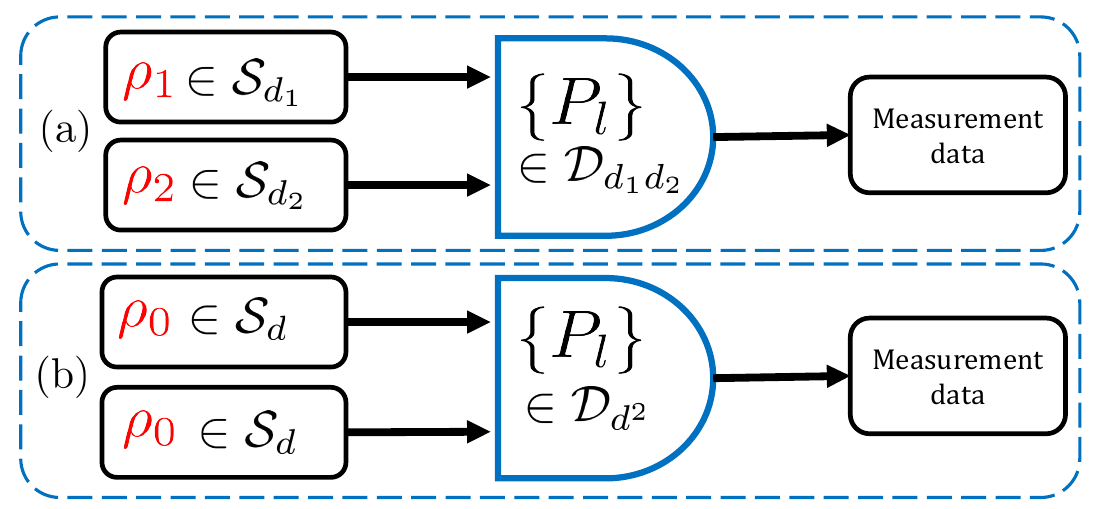}
\centering{\caption{Schematic diagram for two-copy generalized collective QST: (a) The unknown input states are distinct (D-QST) as $\rho_1 \in \mathcal{S}_{d_1}$ and $\rho_2 \in \mathcal{S}_{d_2}$, which may have different dimensions. The measurement operators are $\{P_l\}_{l=1}^L \in \mathcal{D}_{d_1d_2}$.  (b) The unknown input states are identical (I-QST) as $\rho_0\in \mathcal{S}_{d}$ and the measurement operators are $\{P_l\}_{l=1}^L\in \mathcal{D}_{d^2}$.
		}\label{collectiveqst}}
\end{figure}

As illustrated in Fig. \ref{collectiveqst}, we consider a two-copy generalized collective  QST, which includes two different scenarios. Case (a) involves distinct unknown states, labeled as $\rho_1 \in \mathcal{S}_{d_1}$ and $\rho_2 \in \mathcal{S}_{d_2}$, collectively measured by $\{P_l\}_{l=1}^{L} \in \mathcal{D}_{d_1d_2}$ on the large space,
 and is referred to as D-QST in this paper. Case (b) considers two identical copies of a state $\rho_{0}$, referred to 
here as I-QST, which corresponds to the more commonly studied setting in 
the collective QST literature~\cite{Hou2018,conlon2023}.
 Since we consider both cases, we term the approach generalized collective QST.

For D-QST, we assume their dimensions $d_1$ and $d_2$ can be different. Let the complete basis set be $\{\Omega_{i}\}_{i=1}^{d_1^2} $ and $\{\Xi_{j}\}_{j=1}^{d_2^2}$, respectively. Then
we  expand $ \rho_1 $, $\rho_2$ into these bases  as
\begin{equation}\label{rho1}
	\begin{aligned}
	\rho_1 =&\sum_{i=1}^{d_1^2} 	\theta_{1}^{i}\Omega_i, \quad \rho_2 =\sum_{j=1}^{d_2^2} 	\theta_{2}^j\Xi_j,\\
		\theta_1\triangleq&[\theta_{1}^{1}, \theta_{1}^{2}, \cdots, \theta_{1}^{d_1^2}]^{T}, \; \theta_2\triangleq[\theta_{2}^{1}, \theta_{2}^{2}, \cdots, \theta_{2}^{d_2^2}]^{T},
		\end{aligned}
\end{equation}
where $\theta_1 \in \mathbb{R}^{d_1^2}$ and $\theta_2 \in \mathbb{R}^{d_2^2}$.
We also denote the inverse map from $\theta_i$ to $\rho_i$ $(i=1, 2)$ as $h_{i}(\cdot):  \mathbb{R}^{d_i^2}\rightarrow \mathbb{C}^{d_i\times d_i}$.
Let the POVM elements be $\{P_l\}_{l=1}^{L}$ and $P_l \in \mathbb C^{d_1d_2\times d_1d_2}$.
Using the complete orthonormal basis $\{\Omega_{i}\}_{i=1}^{d_1^2} \otimes \{\Xi_{j}\}_{j=1}^{d_2^2} $, we can expand $ P_l$ as
\begin{equation}
	\begin{aligned}
	P_l =&\sum_{i,j=1}^{d_1^2,d_2^2} \phi_{l}^{i,j} (\Omega_i \otimes \Xi_j),  \\
	\phi_{l}\triangleq&\Big[\phi_{l}^{1,1}, \phi_{l}^{1,2},  \cdots, \phi_{l}^{d_1^2,d_2^2}\Big]^{T},
		\end{aligned}
\end{equation}
where $\phi_{l} \in \mathbb{R}^{d_1^2 d_2^2}$.
Therefore, the ideal measurement results of $P_l$ on the state $\rho_1$, $\rho_2$ is 
\begin{equation}
	\begin{aligned}
		p_{l}&=\operatorname{Tr}\left((\rho_1 \otimes \rho_2) 	P_l \right)\\
		&= \phi_{l}^{T} (\theta_1 \otimes \theta_2).
	\end{aligned}
\end{equation}
Define 
\begin{equation}
	\Phi \triangleq [\phi_{1}, \cdots, \phi_{L} ]^{T}, \quad Y\triangleq[p_1, \cdots, p_L]^{T},
\end{equation}
and the problem to identify $\rho_{1}$ and $\rho_{2}$ with two-copy collective POVM can be formulated as follows:
\begin{problem}\label{p1}
	Given the matrix $ \Phi$ and experimental data $\hat Y$, solve $\min_{\theta_1, \theta_2 }|| \hat Y-\Phi\left(\theta_1\otimes \theta_2\right)||^2$ such that $h_1(\theta_1) \in \mathcal{S}_{d_1}$ and $h_2(\theta_2) \in \mathcal{S}_{d_2}$.
\end{problem}

For I-QST, we  expand $\rho_{0} \in \mathcal{S}_{d}$ in the basis $\{\Omega_{i}\}_{i=1}^{d^2} $ as
\begin{equation}\label{rho0}
	\begin{aligned}
		\rho_0 =&\sum_{i=1}^{d^2} 	\theta_{0,i}\Omega_i, \\
		\theta_0\triangleq&[\theta_{0,1}, \theta_{0,2}, \cdots, \theta_{0,d^2}]^{T},
	\end{aligned}
\end{equation}
and define $h_{0}(\cdot):  \mathbb{R}^{d^2}\rightarrow \mathbb{C}^{d\times d}$, where $h_0(\theta_0)=\rho_0$.
We can also  parameterize $\{P_l\}_{l=1}^L \in \mathcal{D}_{d^2}$ to obtain $\Phi$.
Therefore, we can formulate I-QST as the following problem.
\begin{problem}\label{p2}
	Given the matrix $ \Phi$ and experimental data $\hat Y$, solve $\min_{\theta_0 }|| \hat Y-\Phi\left(\theta_0\otimes \theta_0\right)||^2$ such that $h_0(\theta_0)\in  \mathcal{S}_{d}$.
\end{problem}

\begin{remark}\label{re1}
	The two-copy generalized collective QST can also be extended to $n$-copy where the cost functions for Problems \ref{p1} and \ref{p2} become $\|Y-\Phi\left(\theta_1\otimes \cdots \otimes \theta_n\right)\|^2 $ and $\|Y-\Phi\left(\theta_0^{\otimes n} \right)\|^2 $, respectively.
\end{remark}

\subsection{Generalized collective QDT}

\begin{figure}
	\centering
	\includegraphics[width=3.4in]{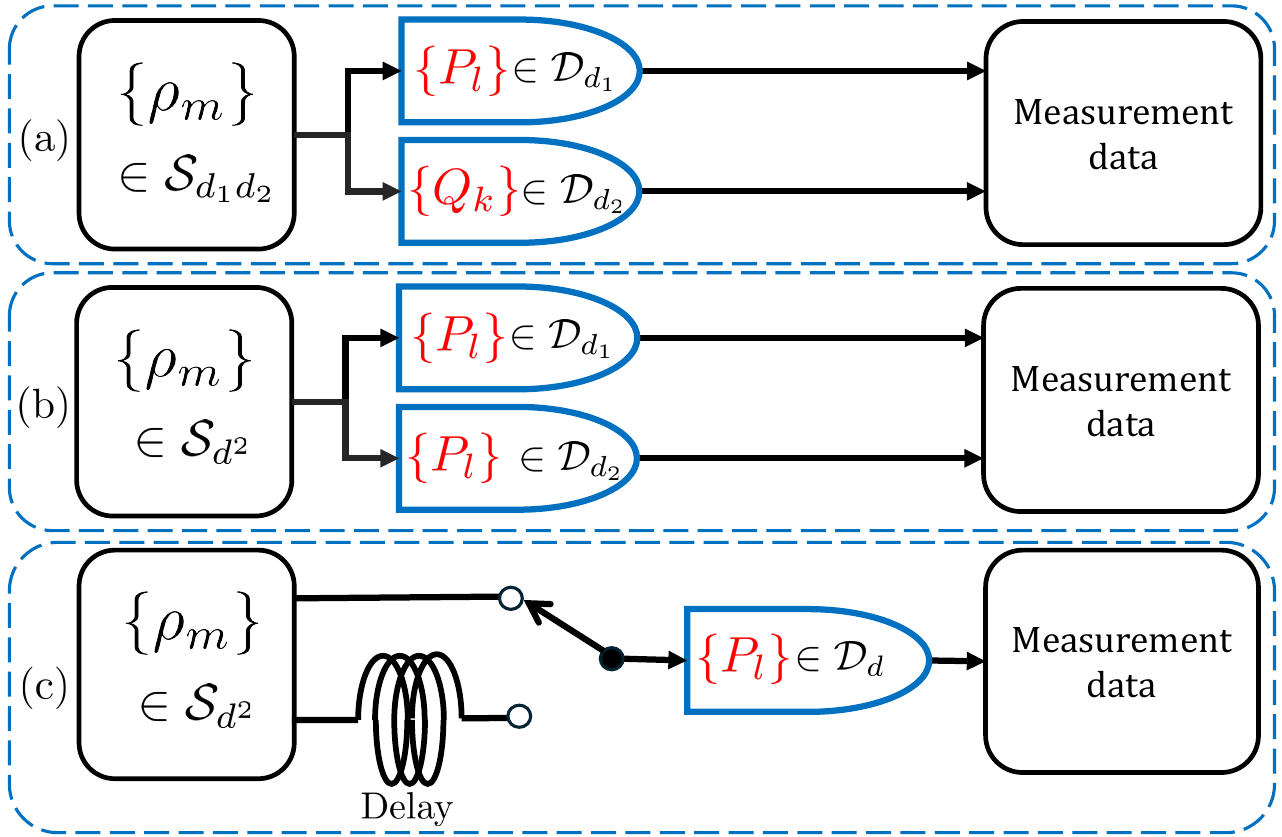}
	\centering{\caption{Schematic diagram for two-copy generalized collective QDT: (a) the unknown detectors are distinct (D-QDT) as $\{P_l\}_{l=1}^{L} \in  \mathcal{D}_{d_1}  $ and $\{Q_k\}_{k=1}^{K} \in \mathcal{D}_{d_2}$, and the probe states are $\{\rho_{m}\}_{m=1}^{M} $ where $\rho_{m} \in \mathcal{S}_{d_1d_2}$. (b) The detectors are identical (I-QDT) as $\{P_l\}_{l=1}^{L} \in  \mathcal{D}_{d}$, and the probe states are $\{\rho_{m}\}_{m=1}^{M} $ where $\rho_{m} \in \mathcal{S}_{d^2}$. This can be realized by, e.g., a fiber delay line \cite{Yokoyama2013,hi2rohtua} and a switch as in (c), where the switch toggled after the measurement of $\operatorname{Tr}_2(\rho_m)$ is completed. }\label{collectiveqdt}}
\end{figure}

We then consider two-copy generalized collective QDT as Fig. \ref{collectiveqdt}, where we also consider two scenarios.
In Case (a), referred to as D-QDT, the unknown detectors are distinct as $\{P_l\}_{l=1}^{L} \in \mathcal{D}_{d_1}$ and $\{Q_k\}_{k=1}^{K} \in \mathcal{D}_{d_2}$ which can be expanded 
in the bases $\{\Omega_{i}\}_{i=1}^{d_1^2}$ and $\{\Xi_{j}\}_{j=1}^{d_2^2}$  as
\begin{equation}\label{p}
		\begin{aligned}
	P_l =&\sum_{i=1}^{d_1^2} 	\phi_{l}^{i}\Omega_i, \quad Q_k =\sum_{j=1}^{d_2^2} 	\varphi_{j}^{k}\Xi_j,	\\
	\phi_l\triangleq&[\phi_{l}^{1}, \phi_l^{2}, \cdots, \phi_{l}^{d_1^2}]^{T}, 	\varphi_k\triangleq[\varphi_{k}^{1}, \varphi_k^{2}, \cdots, \varphi_{k}^{d_2^2}]^{T}.
		\end{aligned}
\end{equation}
Let the probe states be $\{\rho_{m}\}_{m=1}^{M}, \rho_{m} \in \mathcal{S}_{d_1 d_2}$ and we can expand $ \rho_m$ in the basis $\{\Omega_{i}\}_{i=1}^{d_1^2} \otimes \{\Xi_{j}\}_{j=1}^{d_2^2} $ as
\begin{equation}
	\begin{aligned}
	\rho_{m} &=\sum_{i,j=1}^{d_1^2,d_2^2} \theta_{m}^{i,j} (\Omega_i \otimes \Xi_j), \\ \theta_{m}&\triangleq\Big[\theta_{m}^{1,1}, \theta_{m}^{1,2},  \cdots, \theta_{m}^{d_1^2,d_2^2}\Big]^{T}.
	\end{aligned}
\end{equation} 
Therefore, the ideal measurement result of the POVM element $P_l \otimes Q_k$ on  $\rho_m$ is 
\begin{equation}
	\begin{aligned}
		p_{lk}^{m}&=\operatorname{Tr}\left(\rho_{m}(P_l \otimes Q_k) 	 \right)\\
		&= \theta_{m}^{T} (\phi_l \otimes \varphi_k).
	\end{aligned}
\end{equation}
Define 
\begin{equation}
	\Theta \triangleq [\theta_{1}, \cdots, \theta_{M} ]^{T}, \quad Y_{lk}=[p_{lk}^{1}, \cdots, p_{lk}^{M}]^{T},
\end{equation}
and thus
\begin{equation}
	Y_{lk}=\Theta (\phi_l \otimes \varphi_k).
\end{equation}
Therefore, the problem to identify $\{P_l\}_{l=1}^{L}$ and $\{Q_k\}_{k=1}^{K}$ using quantum states can be formulated as follows:
\begin{problem}\label{pd1}
	Given the matrix $ \Theta$ and experimental data $\{\hat Y_{lk}\}$, solve $$\min_{\{\phi_l\}, \{\varphi_k\}} \sum_{l,k=1}^{L,K}|| \hat Y_{lk}-\Theta (\phi_l \otimes \varphi_k)||^2,$$ where $\{h_1(\phi_l)\}_{l=1}^{L} \in \mathcal{D}_{d_1}$, $\{h_2(\varphi_k)\}_{k=1}^{K} \in \mathcal{D}_{d_2}$.
\end{problem}

Case (b), referred to as I-QDT, is also feasible in optical experiments because a fiber delay line can be introduced \cite{Yokoyama2013,hi2rohtua},  allowing us to use the same detector $\{P_l\}_{l=1}^{L} \in \mathcal{D}_{d^2}$ twice sequentially in time by switching between the two paths, as shown in Fig.~\ref{collectiveqdt}(c). A fiber delay line  may introduce losses and imperfections in practice.  While our approach is capable of modeling these effects, for simplicity, we assume them to be negligible in this work.

 Consequently, we can derive the following optimization problem for I-QDT.
\begin{problem}\label{pd2}
	Given the matrix $ \Theta$ and experimental data $\{\hat Y_{lk}\}$, solve $$\min_{\{\phi_l\}} \sum_{l,k=1}^{L,L}|| \hat Y_{lk}-\Theta (\phi_l \otimes \phi_k)||^2$$ where $\{h_0(\phi_l)\}_{l=1}^{L} \in \mathcal{D}_{d}$.
\end{problem}

\begin{remark}\label{re2}
	Two-copy generalized collective QDT can also be extended to $n$-copy collective QDT. For example, for I-QDT, let $\phi_{j_k}^{k}$ be the parameterization vector for the $k$-th qubit of the $j$-th POVM element and the measurement result is $\hat Y_{j_1j_2\cdots j_n}$. The cost function of Problem \ref{pd2}  becomes
	  $\sum_{j_1j_2\cdots j_n}\|\hat Y_{j_1j_2\cdots j_n}-\Theta (\phi_{j_1}^{1} \otimes \phi_{j_2}^{2} \cdots \otimes \phi_{j_n}^{n})\|^2 $.
\end{remark}

\subsection{Generalized collective QPT}

\begin{figure}
	\centering
	\includegraphics[width=3.4in]{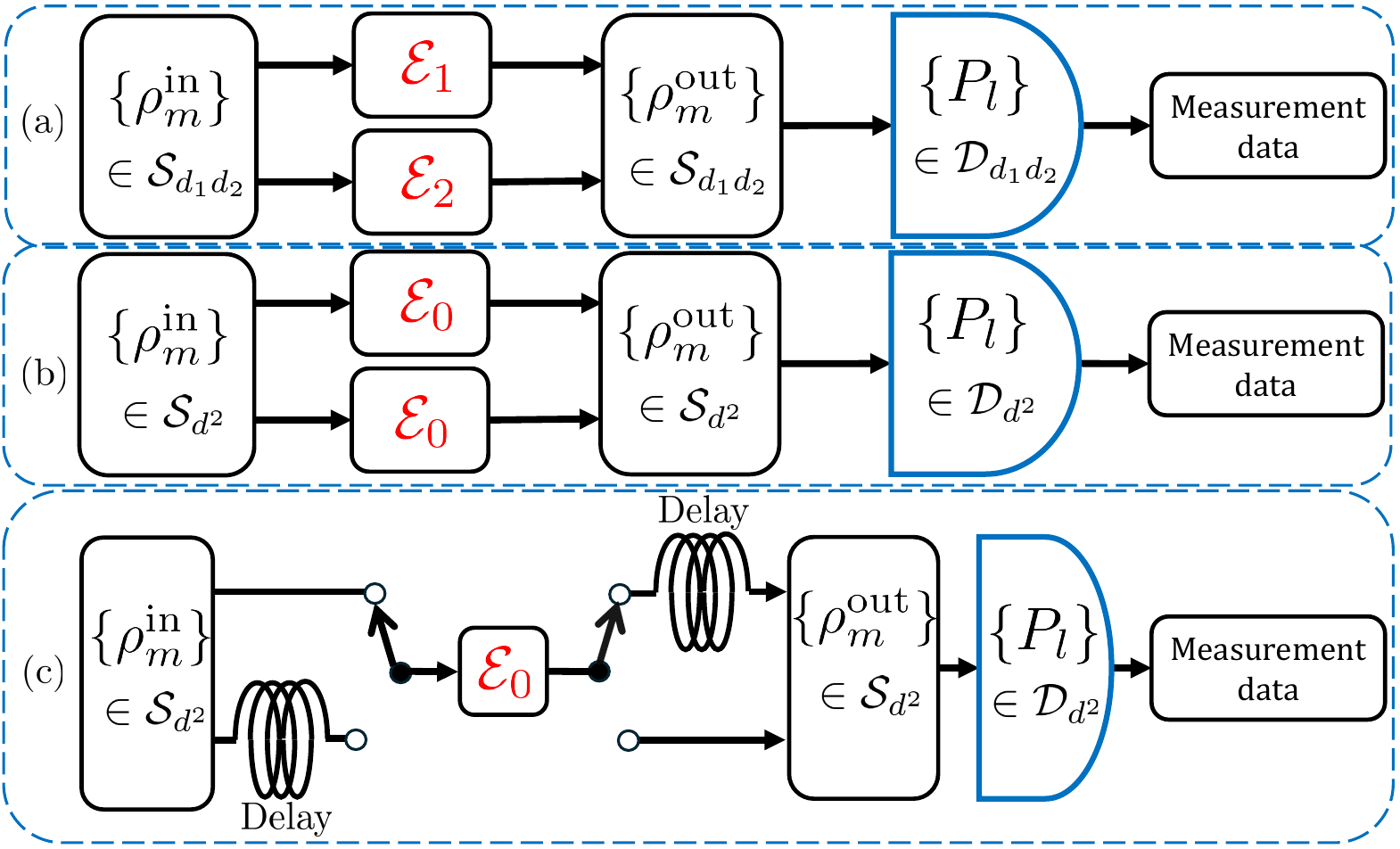}
\centering{\caption{Schematic diagram for two-copy generalized collective QPT: (a) the unknown quantum processes are distinct (D-QPT) as  $\mathcal{E}_1$ and $\mathcal{E}_2$ where the dimensions of the processes are $d_1$ and $d_2$, respectively. The input states are $\{\rho_{m}^{\operatorname{in}}\}_{m=1}^{M}$ where $\rho_{m}^{\operatorname{in}}\in \mathcal{S}_{d_1d_2}$ and the measurement operators are $\{P_l\}_{l=1}^{L} \in \mathcal{D}_{d_1d_2}$. (b) The unknown quantum processes are the same as $\mathcal{E}_0$ (I-QPT) with dimension $d$. The input states are $\{\rho_{m}^{\operatorname{in}}\}_{m=1}^{M} $ where $\rho_{m}^{\operatorname{in}}\in \mathcal{S}_{d^2}$ and the measurement operators are $\{P_l\}_{l=1}^{L} \in \mathcal{D}_{d^2} $. This can be realized by, e.g., two fiber delay lines and two switches as in (c), where the two switches both toggled after $\operatorname{Tr}_2(\rho_m^{\operatorname{in}})$ has passed through $\mathcal{E}_0$.}\label{collectiveqpt}}
\end{figure}

We consider generalized collective QPT as  illustrated in Fig. \ref{collectiveqpt}, where the problem formulation is based on the natural basis $\{|j\rangle\langle k|\}_{1\leq j,k\leq d}$ rather than the basis $\{\Omega_{i}\}_{i=1}^{d^2}$.
We also analyze two cases: in Case (a), referred to as D-QPT, the processes are distinct with dimensions of $\mathcal{E}_1$ and $\mathcal{E}_2$ being $d_1$ and $d_2$, respectively. 
Let the Kraus operators for $ \mathcal{E}_{1}, \mathcal{E}_{2} $ be $ \{  A_{i}^{1}\}_{i=1}^{d_1^2} $ and $ \{  A_{j}^{2}\}_{j=1}^{d_2^2} $, respectively, with $A_{i}^{1} \in \mathbb{C}^{{d_1}\times {d_2}},  A_{j}^{2} \in \mathbb{C}^{{d_2}\times {d_2}}   $. The corresponding  parameterization matrices for  $ \{  A_{i}^{1}\}_{i=1}^{d_1^2} $ and $ \{ A_{j}^{2}\}_{j=1}^{d_2^2} $ are $ C_1 $ and $ C_2 $, respectively, as defined analogously to \eqref{cc}.
The Kraus operators for $ \mathcal{E}=\mathcal{E}_{1}\otimes \mathcal{E}_{2} $ are $ \{  A_{i}^{1} \otimes   A_{j}^{2} \}_{i,j=1}^{d_1^2, d_2^2} $, and the corresponding  parameterization matrix is denoted as $ C $. 
According to  Lemma~\ref{pro1} in Appendix \ref{appendixa}, we have
\begin{equation}
	\begin{aligned}
		\left(\operatorname{vec}\left(  A_{i}^{1} \otimes   A_{j}^{2} \right)\right)^{T}\!\!\!\!=\!\!\left((\operatorname{vec} (  A_{i}^{1}))^{T} \!\otimes\! (\operatorname{vec} ( A_{j}^{2}))^{T}\right)\!\!\left(I_{d_1} \!\otimes\! K_{d_2 d_1}^{T}\!\otimes\! I_{d_2}\right),
	\end{aligned}
\end{equation}
where $  K_{d_2 d_1} $ is a commutation matrix such that $ K_{d_2 d_1} \operatorname{vec}(O)=\operatorname{vec}(O^{T}) $ and $ O $ is a $ d_2 \times d_1$ matrix.
combining it with \eqref{cc}, we have
\begin{equation}\label{cre}
	C=( C_1 \otimes  C_2)\left(I_{d_1} \otimes K_{d_2 d_1}^{T} \otimes I_{d_2}\right).
\end{equation}
With process matrices  $ X_1= C_1^{T}C_1^{*} \in \mathbb{C}^{d_1^2 \times d_1^2}$, $ X_2= C_2^{T}C_2^{*} \in \mathbb{C}^{d_2^2 \times d_2^2} $, and $ X= C^{T}C^{*} \in \mathbb{C}^{d_1^2d_2^2 \times d_1^2d_2^2} $, using \eqref{cre}, we have
\begin{equation}\label{iso}
	X=\left(I_{d_1} \otimes K_{d_2 d_1} \otimes I_{d_2}\right) \left(X_1 \otimes X_2 \right)\left(I_{d_1} \otimes K_{d_2 d_1}^{T} \otimes I_{d_2}\right).
\end{equation}
Here we consider both TP or non-TP processes, i.e., $
X_1 \in \mathcal{P}_{d_1}^{\operatorname{TP}} / \mathcal{P}_{d_1}^{\neg\operatorname{TP}},
X_2 \in \mathcal{P}_{d_2}^{\operatorname{TP}} / \mathcal{P}_{d_2}^{\neg\operatorname{TP}}$ as defined in Table~\ref{tablet1}.

Denoting $\mathbb{K}_{d_1 d_2} \triangleq I_{d_1} \otimes K_{d_2 d_1} \otimes I_{d_2}$, and using \eqref{property2}, \eqref{iso} and Lemma \ref{pro1} in Appendix \ref{appendixa}, we have
\begin{equation}
	\begin{aligned}
	\operatorname{vec}(X)&=(\mathbb{K}_{d_1 d_2} \otimes \mathbb{K}_{d_1 d_2})\operatorname{vec}(X_1\otimes X_2)\\
	&=(\mathbb{K}_{d_1 d_2} \otimes \mathbb{K}_{d_1 d_2})\mathbb{K}_{d_1^2 d_2^2} (\operatorname{vec}(X_1)\otimes (\operatorname{vec}(X_2) ).
		\end{aligned}
\end{equation}
Let the input states be $\{\rho_{m}^{\operatorname{in}}\}_{m=1}^{M}$, where 
$\rho_{m}^{\operatorname{in}} \in \mathcal{S}_{d_1 d_2}$. 
Following \cite{xiao2024,8022944}, we construct the parameterization 
matrix $B \in \mathbb{C}^{M d_1^2 d_2^2 \times d_1^4 d_2^4}$ using all the input 
states, and define $Y \in \mathbb{C}^{M d_1^2 d_2^2}$ as the vector obtained 
from all the reconstructed output states. The relationship among the input 
states, output states, and the process can then be expressed as a linear equation 
$B \operatorname{vec}(X) = Y$ \cite{xiao2024,8022944}, which further implies
\begin{equation}
   B (\mathbb{K}_{d_1 d_2} \otimes \mathbb{K}_{d_1 d_2}) \mathbb{K}_{d_1^2 d_2^2} 
   \bigl(\operatorname{vec}(X_1) \otimes \operatorname{vec}(X_2)\bigr) = Y.
\end{equation}
Let $\mathcal{B}\triangleq B(\mathbb{K}_{d_2 d_1} \otimes \mathbb{K}_{d_1 d_2})\mathbb{K}_{d_1^2 d_2^2}$.
Therefore, 
 the problem of two-copy generalized collective QPT can be formulated as follows:
 \begin{problem}\label{qpt}
 	Given the matrix $ \mathcal{B}$ and experimental data $\hat Y$, solve $$\min_{X_1, X_2} \big\| \hat Y-\mathcal{B}(\operatorname{vec}(X_1)\otimes \operatorname{vec}(X_2))\big\|^2,$$
 where $X_1 \in \mathcal{P}_{d_1}^{\operatorname{TP}}/\mathcal{P}_{d_1}^{\neg\operatorname{TP}}$ and $X_2 \in\mathcal{P}_{d_2}^{\operatorname{TP}}/\mathcal{P}_{d_2}^{\neg\operatorname{TP}}$ are the process matrices for $\mathcal{E}_1$ and $\mathcal{E}_2$, respectively.
 \end{problem}

 In Case (b), referred to as I-QPT in Fig \ref{collectiveqpt}, the process matrices are both $X_0$ and thus
the cost function simplifies to $\min_{X_0} \big\| \hat Y-\mathcal{B}(\operatorname{vec}(X_0)\otimes \operatorname{vec}(X_0))\big\|^2$ where $X_0 \in \mathcal{P}_{d}^{\operatorname{TP}}/\mathcal{P}_{d}^{\neg\operatorname{TP}}$. Case (b) can be realized via, e.g., two delay lines and two switches as shown in Fig.~\ref{collectiveqpt}(c). Moreover, we can  also extend the two-copy collective QPT to multiple-copy collective QPT, as in Remarks \ref{re1} and \ref{re2}.

\begin{remark}
For Case (a) in Fig. \ref{collectiveqpt}, if $\mathcal{E}_2 = \mathcal{I}_{d_2}$ (identity process), the framework reduces to Ancilla-Assisted Process Tomography (AAPT) \cite{qci,PhysRevA.63.020101,xiaocdc2}. A key advantage of AAPT is that only a single input state with a full Schmidt number is required to reconstruct the unknown process $\mathcal{E}_1$. However, in our case, we assume both $\mathcal{E}_1$ and $\mathcal{E}_2$ are unknown and distinct, making the framework more general than AAPT. The price of this generalization is that it requires a larger set of input states.
\end{remark}

\subsection{Weakly informational-complete scenario}
In QST,  measurements are \emph{informationally complete} if their outcome statistics uniquely determine the quantum state~\cite{sic,prugovevcki1977}. This property can be verified by checking the rank of the parameterization matrix of the measurement operators~\cite{MU2020108837,sic}. In collective tomography problems, we generalize this notion as \emph{informationally complete scenarios}, where the measurement outcome statistics uniquely determine the unknown quantum states, detectors, or processes. Since a rigorous definition is challenging, tied to the uniqueness of solutions to a nonconvex problem, we adopt a relaxed criterion for our generalized collective tomography as follows.
\begin{definition}\label{def1}
	The corresponding scenario is \emph{weakly informational-complete}  if $\operatorname{rank}({\Phi})=d_1^2 d_2^2$ in Problem \ref{p1}, $\operatorname{rank}({\Phi})=d^4$ in Problem \ref{p2}, $\operatorname{rank}(\Theta)=d_1^2d_2^2$ in Problem \ref{pd1}, $\operatorname{rank}(\Theta)=d^4$ in Problem \ref{pd2}, $\operatorname{rank}(\mathcal{B})=d_1^4 d_2^4$  in Problem \ref{qpt} , or $\operatorname{rank}(\mathcal{B})=d^8$ for I-QPT.
\end{definition}

Note that for QPT, we have $\operatorname{rank}(\mathcal{B})=\operatorname{rank}({B})$ because $(\mathbb{K}_{d_1 d_2} \otimes \mathbb{K}_{d_1 d_2})\mathbb{K}_{d_1^2 d_2^2}$  is full-rank.
Conversely, when the rank of the linear map is smaller than the corresponding limit in Definition~\ref{def1}, we denote it as the \emph{weakly informational-incomplete} scenario. 
These rank-based criteria are easy to verify, and weakly informational-completeness guarantees a unique closed-form solution, though it is only a sufficient condition for informational completeness and may be more conservative than those in prior works on single-copy quantum tomography \cite{Qi2013,wang2019twostage,xiao2024, zhu}.

\section{Closed-form solutions}\label{closed}
In this section, we present closed-form solutions for all two-copy generalized collective quantum tomography. We will describe the design of the algorithms and provide an analysis of the computational complexities and MSE scalings.
Additionally, we demonstrate how the algorithms can be extended to three or more-copy scenarios while maintaining the same MSE scalings.

\subsection{Algorithm design}\label{sec31}

\subsubsection{D-QST and I-QST}
For the two-copy generalized collective QST in Problem \ref{p1} of D-QST, we decompose it into two subproblems. The first subproblem is formulated as follows:
\renewcommand{\theproblem}{\arabic{problem}}
\addtocounter{problem}{-5}
\renewcommand{\theproblem}{\arabic{problem}{.1}}
\begin{problem}\label{subproblem1}
Given the matrix $ \Phi$ and experimental data $\hat Y$, solve $\min_{x }|| \hat Y-\Phi x||^2$ where $x\in \mathbb{R}^{d_1^2 d_2^2}$.
\end{problem}

For Problem \ref{subproblem1} when $\Phi$ is full-rank,  we can obtain a unique optimal estimate $\tilde x$ using the least squares method
\begin{equation}\label{s1}
\tilde x=(\Phi^T \Phi)^{-1}\Phi^{T}\hat Y.
\end{equation}
When $\Phi$ is rank-deficient, the data is insufficient to uniquely determine the unknown. One possible reconstruction is given by 
\begin{equation}\label{ss1}
\tilde  x=\Phi^{+}\hat{Y},
\end{equation}
where $\Phi^{+}$ denotes the Moore–Penrose (MP) inverse  of $\Phi$~\cite{horn_johnson_2012}.
Alternatively, we can incorporate a regularization term in the cost function as $||\hat Y-\Phi x||^2+x^{T} D x$
where $D\geq 0$ is a regularization matrix. The closed-form estimate with regularization is given by
\begin{equation}\label{re}
\tilde  x=(\Phi^T\Phi +D)^{-1}\Phi^{T}\hat Y.
\end{equation}
The question of designing the regularization matrix $D$ and corresponding hyperparameters in $D$ has been discussed in~\cite{6883125,Chen2018,Pillonetto2014,10273596}.

After obtaining $\tilde x$, the corresponding $\tilde \rho$ is
\begin{equation}
	\tilde \rho=\sum_{i,j=1}^{d_1^2,d_2^2}  \tilde  x_{(i-1)d_2^2+j} (\Omega_i \otimes \Xi_j).
\end{equation}
Alternatively, we can incorporate the unit trace constraint directly into the parameterization of $x$.
In the complete orthonormal basis $\{\Omega_{i}\}_{i=1}^{d_1^2} \otimes \{\Xi_{j}\}_{j=1}^{d_2^2} $, if one takes, for example, $\Omega_{1}=\frac{I_{d_1}}{\sqrt{d_1}}$ and $\Xi_{1}=\frac{I_{d_2}}{\sqrt{d_2}}$, then
the first element of $x$ should be $\frac{1}{\sqrt{d_1 d_2}}$.  Let $$x=\left[\frac{1}{\sqrt{d_1 d_2}}, x_b^{T}\right]^{T}$$ where $x_b \in \mathbb{R}^{d_1^2d_2^2-1}$, and we partition $\Phi$  as
\begin{equation}
	\Phi=\begin{bmatrix}
		\Phi_a, \; \Phi_{b}
	\end{bmatrix},
\end{equation}
where $\Phi_a \in \mathbb{R}^{L}$ and $\Phi_b \in \mathbb{R}^{L\times (d_1^2d_2^2-1)}$.
Therefore, we have
\begin{equation}
	\Phi_b x_b =Y-\frac{1}{\sqrt{d_1 d_2}}\Phi_a,
\end{equation}
and the least squares solution in the weakly informational-complete scenario is given by
\begin{equation}\label{x1}
	\tilde  x_b= (\Phi_b^T \Phi_b)^{-1}\Phi_b^{T}\Big(\hat Y-\frac{1}{\sqrt{d_1 d_2}}\Phi_a\Big).
\end{equation}
Similar to \eqref{re}, we can also obtain a closed-form solution with a regularization term. Using \eqref{x1}, we reconstruct $\tilde\rho$ as
\begin{equation}\label{bar2}
	\tilde \rho=\frac{1}{{d_1 d_2}}I_{d_1d_2}+\sum_{(i,j)\neq(1,1)}^{d_1^2,d_2^2}  \tilde  x_{(i-1)d_2^2+j} (\Omega_i \otimes \Xi_j).
\end{equation}

We then consider the second sub-problem using $\tilde \rho$.

\renewcommand{\theproblem}{\arabic{problem}}
\addtocounter{problem}{-1}
\renewcommand{\theproblem}{\arabic{problem}{.2}}
\begin{problem}\label{subproblem2}
	Let $\tilde\rho \in \mathbb{C}^{d_1d_2 \times d_1d_2}$ be given.  Solve $\min _{\tilde \rho_1, \tilde \rho_2} \|\tilde \rho-\tilde\rho_1 \otimes \tilde \rho_2\|$ where $\tilde \rho_1 \in \mathcal{S}_{d_1} $, $\tilde \rho_2 \in \mathcal{S}_{d_2}$.
\end{problem}
\renewcommand{\theproblem}{\arabic{problem}}

We can rearrange the elements of $\tilde{\rho}$ and define a permuted version  as $\mathcal{R}(\tilde{\rho}) \in \mathbb{C}^{d_1^2 \times d_2^2}$  \cite{LOAN200085} such that $\mathcal{R}(\tilde\rho_1\otimes \tilde\rho_2)=\operatorname{vec}(\tilde\rho_1)\operatorname{vec}(\tilde\rho_2)^T$. We thus know
\begin{equation}\label{sq2}
\left\|\tilde \rho-\tilde \rho_1 \otimes \tilde \rho_2 \right\|=	\|	\mathcal{R}(\tilde{\rho})-\operatorname{vec}\left(\tilde{\rho}_{1}\right)\operatorname{vec}\left(\tilde{\rho}_{2}\right)^{T}\|.
\end{equation}
Problem \ref{subproblem2} is a nearest Kronecker product problem~\cite{LOAN200085,VanLoan1993}. Using \eqref{sq2},  Problem \ref{subproblem2} can be solved efficiently by using the singular value decomposition (SVD) because it is equivalent to finding the nearest rank-$1$ matrix to a given matrix \cite{LOAN200085,VanLoan1993}. 
If $\tilde{\rho}_1$ and $\tilde{\rho}_2$ are the solutions, then $\alpha \hat{\rho}_1$ and $\frac{1}{\alpha} \hat{\rho}_2$ for arbitrary $\alpha\in \mathbb{C}$  also denotes a solution. Thus, we can choose a proper 
$\alpha \in \mathbb{C}$ such that $\operatorname{Tr}(\tilde{\rho}_1)=1$.
Further considering the Hermitian constraints on $\tilde{\rho}_1, \tilde{\rho}_2$ and the unit trace constraint on $\tilde{\rho}_2$, we take
\begin{equation}
	\bar\rho_1=\frac{{\tilde \rho_1 +\tilde \rho_1^{\dagger}}}{2}, \;	\bar\rho_2=\frac{1}{2}\left(\frac{{\tilde \rho_2 }}{\operatorname{Tr}({\tilde \rho_2})}+\frac{{\tilde \rho_2^{\dagger} }}{\operatorname{Tr}({\tilde \rho_2^{\dagger}})}\right),
\end{equation}
where $\operatorname{Tr}(\bar\rho_1)=\frac{\operatorname{Tr}({\tilde \rho_1 +\tilde \rho_1^{\dagger}})}{2}=1$ because $\operatorname{Tr}(\tilde{\rho}_1)=\operatorname{Tr}(\tilde{\rho}_1^{\dagger})=1$.
However, $\bar{\rho}_1, \bar{\rho}_2$  may not satisfy the positive semidefinite constraint. To address this, we apply the fast correction algorithm from \cite{effqst} to the eigenvalues of $\bar{\rho}$ and obtain the final estimate $\hat{\rho}_1, \hat{\rho}_2  \geq 0$ for Problem \ref{p1}.

For Problem \ref{p2} of I-QST, after the same procedure as solving Problem \ref{subproblem1} and Problem \ref{subproblem2},  we let
\begin{equation}
	\hat\rho_0=\frac{\hat \rho_1 + \hat \rho_2}{2},
\end{equation}
which is a physical estimate because $\hat \rho_1$, $\hat \rho_2 \geq 0$ and $\operatorname{Tr}(\hat \rho_1)=\operatorname{Tr}(\hat \rho_2)=1$.

\begin{figure}[!t]
	\centering
	\includegraphics[width=3.4in]{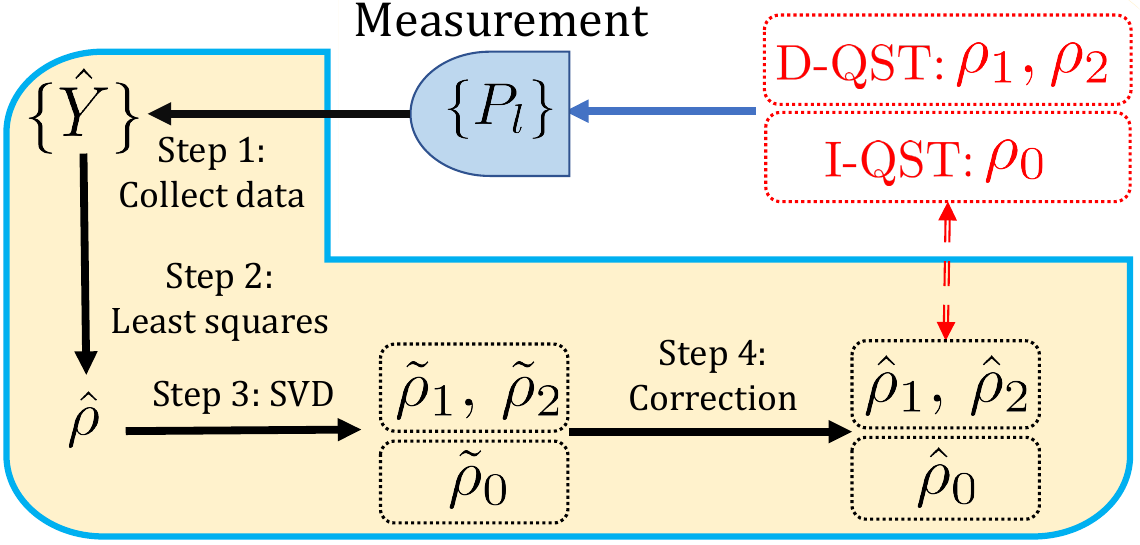}
	\centering{\caption{Steps in our closed-form algorithm for the generalized collective QST.}\label{fst}}
\end{figure}
Overall, the procedures in our closed-form algorithm consist of four steps, as outlined in Fig.~\ref{fst}. The pseudo-code for the closed-form algorithm for D-QST is provided in Algorithm~\ref{A1}.  
Similar to \cite{Qi2013}, we can derive the total online computational complexity as  $O\left(L d_1^2 d_2^2 + d_1^2 d_2^2 \min(d_1^2, d_2^2)\right)$
for D-QST, and  $O(L d^4 + d^6)$ for I-QST. The complexity is dominated by the least-squares method in Step~2 and the SVD in Step~3 of Fig.~\ref{fst} for generalized collective QST. These computational complexities are of the same order as those of typical individual QST with dimension $d_1 d_2$ or $d^2$ \cite{Qi2013}.
\begin{remark}\label{rem4}
In Problem~\ref{subproblem2}, we adopt the Frobenius norm. Alternatively, one may consider other metrics such as the quantum relative entropy $D\left(\cdot\| \cdot\right)$, as discussed in \cite{wilde2013quantum}. In this case, the minimization problem takes the form  
\begin{equation}
   \min_{\tilde \rho_1,\,\tilde \rho_2} 
   D\left(\tilde \rho \| \tilde \rho_1 \otimes \tilde \rho_2\right),
\end{equation}
where the optimal solution is given by the marginals  
$\tilde \rho_1=\operatorname{Tr}_{2}(\tilde\rho)$ and $\tilde \rho_2=\operatorname{Tr}_{1}(\tilde\rho)$.  
The corresponding minimum value of quantum relative entropy coincides with the quantum mutual information between the two subsystems in $\tilde{\rho}$~\cite{wilde2013quantum}.
\end{remark}

\begin{algorithm}[!t]	
	\caption{Closed-form Algorithm for QST (Problem~\ref{p1})}\label{A1}
	\begin{algorithmic}[1]
		\Input{Measurement data $\hat Y$; parametrization matrix $\Phi$ of different quantum measurements}
		\Output{Estimated states $\hat \rho_1$, $\hat \rho_2$}
		
		\If{$\operatorname{rank}(\Phi)=d_1^2 d_2^2$} \Comment{Weakly informational-complete case}
			\State $\tilde x = (\Phi^T \Phi)^{-1}\Phi^T \hat Y$ \Comment{Least-squares solution of Problem~\ref{subproblem1}}
		\Else \Comment{Weakly informational-incomplete case}
			\State $\tilde x = \Phi^{+}\hat Y$ \quad or \quad $\tilde x=(\Phi^T\Phi+D)^{-1}\Phi^T \hat Y$\Comment{Moore–Penrose inverse or regularized solution of Problem~\ref{subproblem1}}
		\EndIf
		
		\State Reconstruct $\tilde \rho$ from $\tilde x$ and apply the permutation $\mathcal{R}(\tilde \rho)$
		\State Solve 
		\[
		\min_{\tilde \rho_1, \tilde \rho_2} \big\| \mathcal{R}(\tilde \rho) - \operatorname{vec}(\tilde \rho_1)\operatorname{vec}(\tilde \rho_2)^T \big\|
		\]
		using SVD, with the normalization $\operatorname{Tr}(\tilde \rho_1)=1$\Comment{Equivalent to Problem~\ref{subproblem2}}
		
		\State Symmetrize and normalize:
		\[
		\bar \rho_1 = \tfrac{1}{2}(\tilde \rho_1 + \tilde \rho_1^{\dagger}), \quad
		\bar \rho_2 = \tfrac{1}{2}\!\left(\tfrac{\tilde \rho_2}{\operatorname{Tr}(\tilde \rho_2)} + \tfrac{\tilde \rho_2^{\dagger}}{\operatorname{Tr}(\tilde \rho_2^{\dagger})}\right)
		\]
		
		\State Apply the correction algorithm of \cite{effqst} to $\bar \rho_1, \bar \rho_2$ to obtain the final estimates $\hat \rho_1, \hat \rho_2$
	\end{algorithmic}
\end{algorithm}

	
		

	

\subsubsection{D-QDT and I-QDT}

We then consider a closed-form solution for generalized collective QDT.
We also split Problem~\ref{pd1} into to two subproblems. The first subproblem is similar to Problem~\ref{subproblem1}, where  the least squares estimate in the weakly informational-complete scenario is
\begin{equation}\label{phi}
	\tilde  \phi_{lk}=(\Theta^T\Theta)^{-1}\Theta^{T}\hat Y_{lk}.
\end{equation}
We can also utilize MP inverse or regularization in the weakly informational-incomplete scenario, which is similar to \eqref{ss1} and \eqref{re}. Then we 
 reconstruct $\tilde{R}_{lk}$ as
\begin{equation}\label{dtf}
\tilde {R}_{lk}=\sum_{i,j=1}^{d_1^2,d_2^2}  (\tilde  \phi_{lk})_{(i-1)d_2^2+j} (\Omega_i \otimes \Xi_j).
\end{equation}
We ultimately derive a total of $KL$ distinct $\tilde {R}_{lk}$ because $1\leq l \leq L, 1\leq k \leq K$. 

We then consider the second subproblem using $\{\tilde{R}_{lk}\}$ as follows:
\renewcommand{\theproblem}{\arabic{problem}{.2}}
\addtocounter{problem}{1}
\begin{problem}\label{subproblem5}
	Let $\tilde  R_{lk} \in \mathbb{C}^{d_1d_2 \times d_1d_2} $ be given.  Solve $\min_{\{\tilde P_l\}, \{\tilde Q_k\}} \sum_{l,k=1}^{L,K}\|\tilde R_{lk}-\tilde P_l \otimes \tilde Q_k \|^2$ where $\{\tilde P_l\} \in \mathcal{D}_{d_1}$ and $\{\tilde Q_k\} \in \mathcal{D}_{d_2}$.
\end{problem}

Define a matrix collecting all of $\{\mathcal{R}(R_{lk})\}$ as follows:
\begin{equation}\label{rr11}
    \tilde R=\begin{bmatrix}
        \mathcal{R}(\tilde R_{11}) & \mathcal{R}(\tilde R_{12}) & \cdots & \mathcal{R}(\tilde R_{1K}) \\
        \vdots & \vdots & \ddots & \vdots \\
        \mathcal{R}(\tilde R_{L1}) & \mathcal{R}(\tilde R_{L2}) & \cdots & \mathcal{\tilde R}(R_{LK})
    \end{bmatrix},
\end{equation}
where $\tilde R \in \mathbb{C}^{L d_1^2 \times K d_2^2}$.  
Define the vectors of all $\tilde P_l$ and $\tilde Q_k$ as
\begin{equation}\label{rr22}
\begin{aligned}
 \operatorname{vec}(\tilde P) &= \big[\operatorname{vec}(\tilde P_1)^{T}, \cdots, \operatorname{vec}(\tilde P_L)^{T}\big]^{T}, \\
 \operatorname{vec}(\tilde Q) &= \big[\operatorname{vec}(\tilde Q_1)^{T}, \cdots, \operatorname{vec}(\tilde Q_K)^{T}\big]^{T},
\end{aligned}
\end{equation}  
and thus
\begin{equation}
\sum_{l,k=1}^{L,K}\|\tilde R_{lk}-\tilde P_l \otimes \tilde Q_k \|^2
=\|\tilde R- \operatorname{vec}(\tilde P)\operatorname{vec}(\tilde Q)^{T}\|^2.
\end{equation}
Therefore, similar to Problem~\ref{subproblem2}, this can be solved by using the SVD. However, note that $\alpha \tilde P$ and $\frac{1}{\alpha} \tilde Q$ (for any $\alpha \in \mathbb{C}$) are also solutions. Since $\operatorname{Tr}\!\left(\sum_{l=1}^{L} P_l\right) = d_1$, we can fix $\alpha$ such that $\operatorname{Tr}\!\left(\sum_{l=1}^{L} \tilde P_l\right) = d$, which then determines the corresponding $\{\tilde Q_k\}_{k=1}^{K}$.  

To satisfy the Hermitian constraint, we define
\begin{equation}
\bar{P}_l = \frac{\tilde{P}_l + \tilde{P}_l^{\dagger}}{2}, \qquad 
\bar{Q}_k = \frac{\tilde{Q}_k + \tilde{Q}_k^{\dagger}}{2}.
\end{equation}
However, these estimates may not satisfy the completeness and positive semidefinite constraints. To enforce these, we apply the Stage-2 algorithm of \cite{wang2019twostage}, yielding the final estimates $\{\hat{P}_l\}_{l=1}^{L}$ and $\{\hat{Q}_k\}_{k=1}^{K}$.

	

	
\begin{algorithm}[!t]	
	\caption{Closed-form Algorithm for QDT (Problem~\ref{pd1})}\label{A3}
	\begin{algorithmic}[1]
		\Input{Measurement data $\{\hat Y_{lk}\}$; parametrization matrix $\Theta$ of different quantum states}
		\Output{Estimated operators $\{\hat P_l\}$, $\{\hat Q_k\}$}
		
		\If{$\operatorname{rank}(\Theta)=d_1^2 d_2^2$} \Comment{Weakly informational-complete case}
			\State $\tilde \phi_{lk} = (\Theta^T \Theta)^{-1}\Theta^T \hat Y_{lk}$ \Comment{Least-squares solution}
		\Else \Comment{Weakly informational-incomplete case}
			\State $\tilde \phi_{lk} = \Theta^{+}\hat Y_{lk}$ \quad or \quad $\tilde \phi_{lk} = (\Theta^T \Theta + D)^{-1}\Theta^T \hat Y_{lk}$ \Comment{Moore–Penrose inverse or regularized solution}
		\EndIf
		
		\State Reconstruct $\{\tilde R_{lk}\}$ from $\{\tilde \phi_{lk}\}$; apply the permutation $\mathcal{R}(\tilde R_{lk})$ and form $\tilde R$, $\operatorname{vec}(\tilde P)$, $\operatorname{vec}(\tilde Q)$ in \eqref{rr11} and \eqref{rr22}
		\State Solve
		\[
		  \min_{\tilde P, \tilde Q} \ \big\|\tilde R - \operatorname{vec}(\tilde P)\operatorname{vec}(\tilde Q)^T\big\|^2
		\]
		using SVD, with the constraint $\operatorname{Tr}\!\left(\sum_{l=1}^{L}\tilde P_l\right)=d$ \Comment{Equivalent to Problem~\ref{subproblem5}}
		
		\State Symmetrize: 
		\[
		\bar P_l = \tfrac{1}{2}(\tilde P_l + \tilde P_l^{\dagger}), 
		\quad 
		\bar Q_k = \tfrac{1}{2}(\tilde Q_k + \tilde Q_k^{\dagger})
		\]
		
		\State Apply the Stage-2 algorithm of \cite{wang2019twostage} to $\{\bar P_l\}, \{\bar Q_k\}$ to obtain the final estimates $\{\hat P_l\}_{l=1}^{L}$ and $\{\hat Q_k\}_{k=1}^{K}$
	\end{algorithmic}
\end{algorithm}

For I-QDT, we obtain the following optimization problem using $\tilde R_{ll}$ for the second subproblem as follows:
\addtocounter{problem}{0}
\renewcommand{\theproblem}{\arabic{problem}{.2}}
\begin{problem}\label{subproblem3}
	Let $\tilde R_{lk} \in \mathbb{C}^{d^2 \times d^2}$ be given. Solve
 \[
 \min_{\{\tilde P_l\} \in \mathcal{D}_d} \ \sum_{l,k=1}^{L,L}\|\tilde R_{lk}-\tilde P_l\otimes \tilde P_k\|^2.
 \]
\end{problem}

Similarly, we can define $\tilde R$, $\tilde P$, and $\tilde Q$ as in \eqref{rr11} and \eqref{rr22}, and obtain
\begin{equation}
  \sum_{l,k=1}^{L,L}\|\tilde R_{lk}-\tilde P_l\otimes \tilde P_k\|^2
  = \|\tilde R - \operatorname{vec}(\tilde P)\operatorname{vec}(\tilde P)^{T}\|.
\end{equation}
This formulation differs from Problem~\ref{subproblem5} in that the tensor terms are identical.  

To address this, we first consider a relaxed version in which the terms are allowed to be distinct:
\begin{equation}\label{sq23}
\sum_{l,k=1}^{L,L}\|\tilde R_{lk}-\tilde P_l\otimes \tilde Q_k\|^2
= \|\tilde R - \operatorname{vec}(\tilde P)\operatorname{vec}(\tilde Q)^{T}\|,
\end{equation}
which is equivalent to Problem~\ref{subproblem5}. This problem can also be solved using the SVD~\cite{LOAN200085, VanLoan1993}, yielding $\tilde P_l$ such that $\operatorname{Tr}\!\left(\sum_{l=1}^{L}\tilde P_l\right) = d$.
Following the same procedure as in D-QDT, we then obtain the final estimates $\{\hat{P}_{l}\}_{l=1}^{L}$.

\begin{figure}[!t]
	\centering
	\includegraphics[width=3.4in]{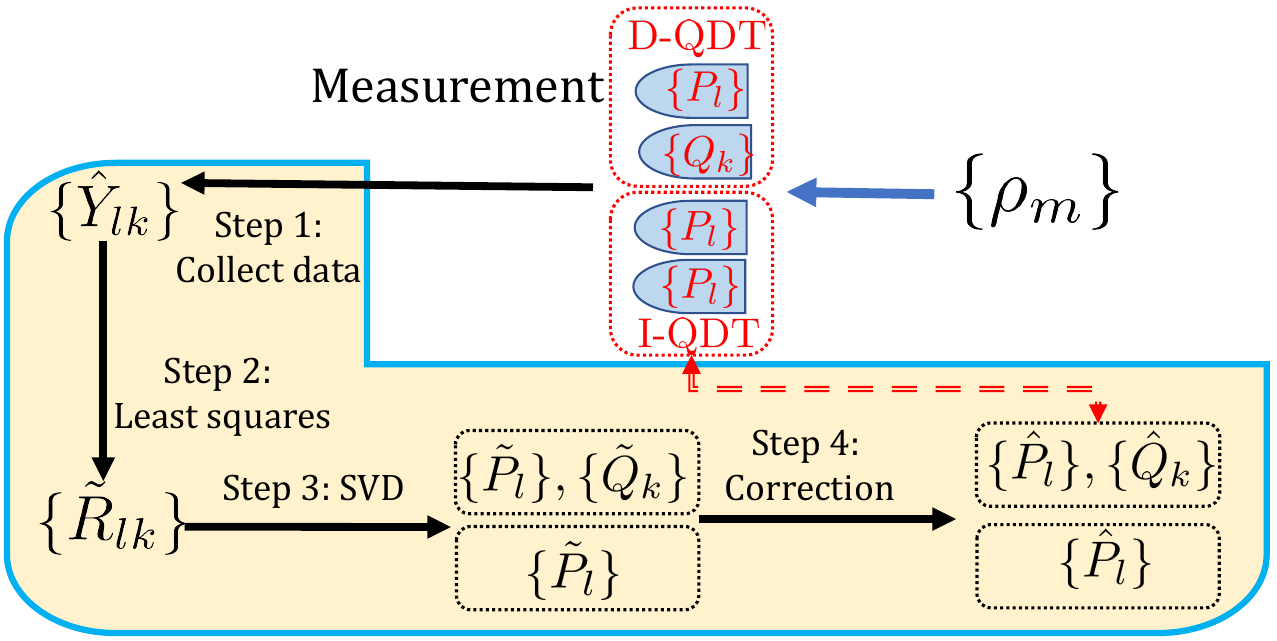}
	\centering{\caption{Steps in our closed-form algorithm for the generalized collective QDT.}\label{fdt}}
\end{figure}


Overall, the procedures of our closed-form algorithm involve four steps, as outlined in Fig.~\ref{fdt}. The pseudo-code of the closed-form algorithm for D-QDT is provided in Algorithm~\ref{A3}. Similar to \cite{wang2019twostage}, the total online computational complexities can be derived as  $O\!\left(MKL d_1^2 d_2^2 + KL d_1^2 d_2^2 \min(L d_1^2, K d_2^2)\right)$
for D-QDT, and $O(M L^2 d^4 + L^3 d^6)$ 
for I-QDT. These complexities are dominated by the least-squares method in Step~2 and the SVD in Step~3 of Fig.~\ref{fdt} for generalized collective QDT. The resulting orders of complexity are comparable to those of typical individual QDT with $KL$ different POVM elements of dimension $d_1 d_2$, or $L^2$ different POVM elements of dimension $d^2$~\cite{wang2019twostage}.

\subsubsection{D-QPT and I-QPT}

We finally  design a closed-form solution for two-copy generalized collective QPT in Problem \ref{qpt}. We first obtain the least squares solution
\begin{equation}
	\tilde  X= \operatorname{vec}^{-1}\left((\mathcal{B}^{\dagger}\mathcal{B})^{-1}\mathcal{B}^{\dagger}\hat Y\right).
\end{equation}
In the weakly informational-incomplete scenario, we can also use the MP inverse or regularization, similar to \eqref{ss1} and \eqref{re}. We then consider the following problem for D-QPT.
\begin{problem}\label{subproblem4}
	Let $\tilde X \in \mathbb{C}^{d_1^2 d_2^2 \times d_1^2 d_2^2}$ be given.  Solve $\min_{\tilde X_1, \tilde X_2} \|\tilde X-\tilde X_1 \otimes \tilde X_2\|$  where $X_i \in \mathcal{P}_{d_i}^{\operatorname{TP}}/\mathcal{P}_{d_i}^{\neg\operatorname{TP}}$.
\end{problem}

If  $\mathcal{E}_i$ $(i=1,2)$ is TP, we have $\operatorname{Tr}(X_i)=d_i$.
 For example, if $\operatorname{Tr}(X_1)=d_1$,
similar to Problem \ref{subproblem2}, we can obtain a unique optimal solution  $\tilde{X}_{1}$ and $\tilde{X}_{2}$ via SVD,  ensuring that $\operatorname{Tr}(\tilde X_1)=d_1$. To correct the solutions so that they satisfy the Hermitian constraint, we let
\begin{equation}\label{eq45}
	\bar X_1=\frac{\tilde{X}_{1}+\tilde{X}_{1}^{\dagger}}{2}, \;\;\bar  X_2=\frac{\tilde{X}_{2}+\tilde{X}_{2}^{\dagger}}{2}.
\end{equation}
If they are both TP, we have $\operatorname{Tr}(X_1)=d_1, \operatorname{Tr}(X_2)=d_2$. After the first equality in \eqref{eq45}, we take
\begin{equation}
	\bar{X}_2=\frac{d_2}{2}\left(\frac{\tilde{X}_2}{\operatorname{Tr}(\tilde{X}_2)}+\frac{\tilde{X}_2^{\dagger}}{\operatorname{Tr}(\tilde{X}_2^{\dagger})}\right).
\end{equation}

 However, if  both $\mathcal{E}_1$ and $\mathcal{E}_2$  are non-TP, the solution to Problem \ref{subproblem4} is not unique because $\beta \tilde X_1$ and  $\frac{1}{\beta} \tilde X_2$ are also solutions. To obtain a unique estimation in this case, we prepare maximally mixed state $\rho=\frac{I_{d_1}}{d_1}$ and implement $P= I_{d_1}$ on the output state $\mathcal{E}_1(\frac{I_{d_1}}{d_1})$, and then obtain a real positive  $\alpha_1$ from measurement result, which is an estimate of $\operatorname{Tr}(X_1)/d_1$. 
 To determine $\beta$, we rescale $\tilde{X}_1$ such that $\operatorname{Tr}(\tilde{X}_1)=d_1\alpha_1$ and obtain the corresponding $\tilde X_2$.
These state and measurement operators are relatively straightforward to generate in an experimental setup.
 Then we obtain the estimates to satisfy Hermitian constraint
\begin{equation}	\bar{X}_1=\frac{\tilde{X}_{1}+\tilde{X}_{1}^{\dagger}}{2}, \;\;\bar{X}_2= \frac{\tilde{X}_{2}+\tilde{X}_{2}^{\dagger}}{2}.
\end{equation}
For I-QPT, similar to I-QDT, we can first obtain $\tilde{X}_0=\frac{\tilde{X}_{1}+\tilde{X}_{2}^{\dagger}}{2}$ and then obtain a unique solution $\bar{X}_0$ in the TP case as
\begin{equation}
	\bar{X}_0=\frac{d}{2}\left(\frac{\tilde{X}_0}{\operatorname{Tr}(\tilde{X}_0)}+\frac{\tilde{X}_0^{\dagger}}{\operatorname{Tr}(\tilde{X}_0^{\dagger})}\right),
\end{equation}
and in the non-TP case as
\begin{equation}
\bar{X}_0=\frac{1}{2}\sqrt{\frac{\operatorname{Tr}(\tilde  X+\tilde  X^{\dagger})}{2}} \left(\frac{\tilde{X}_0}{\operatorname{Tr}(\tilde{X}_0)}+\frac{\tilde{X}_0^{\dagger}}{\operatorname{Tr}(\tilde{X}_0^{\dagger})}\right).
\end{equation}

Finally, to satisfy the partial trace constraint for TP or non-TP process,   we implement the algorithm in \cite{xiao2024} and obtain the final estimate $\hat X_1$ and $\hat X_2$ for D-QPT or $\hat X_0$ for I-QPT.

\begin{figure}[!t]
	\centering
	\includegraphics[width=3.4in]{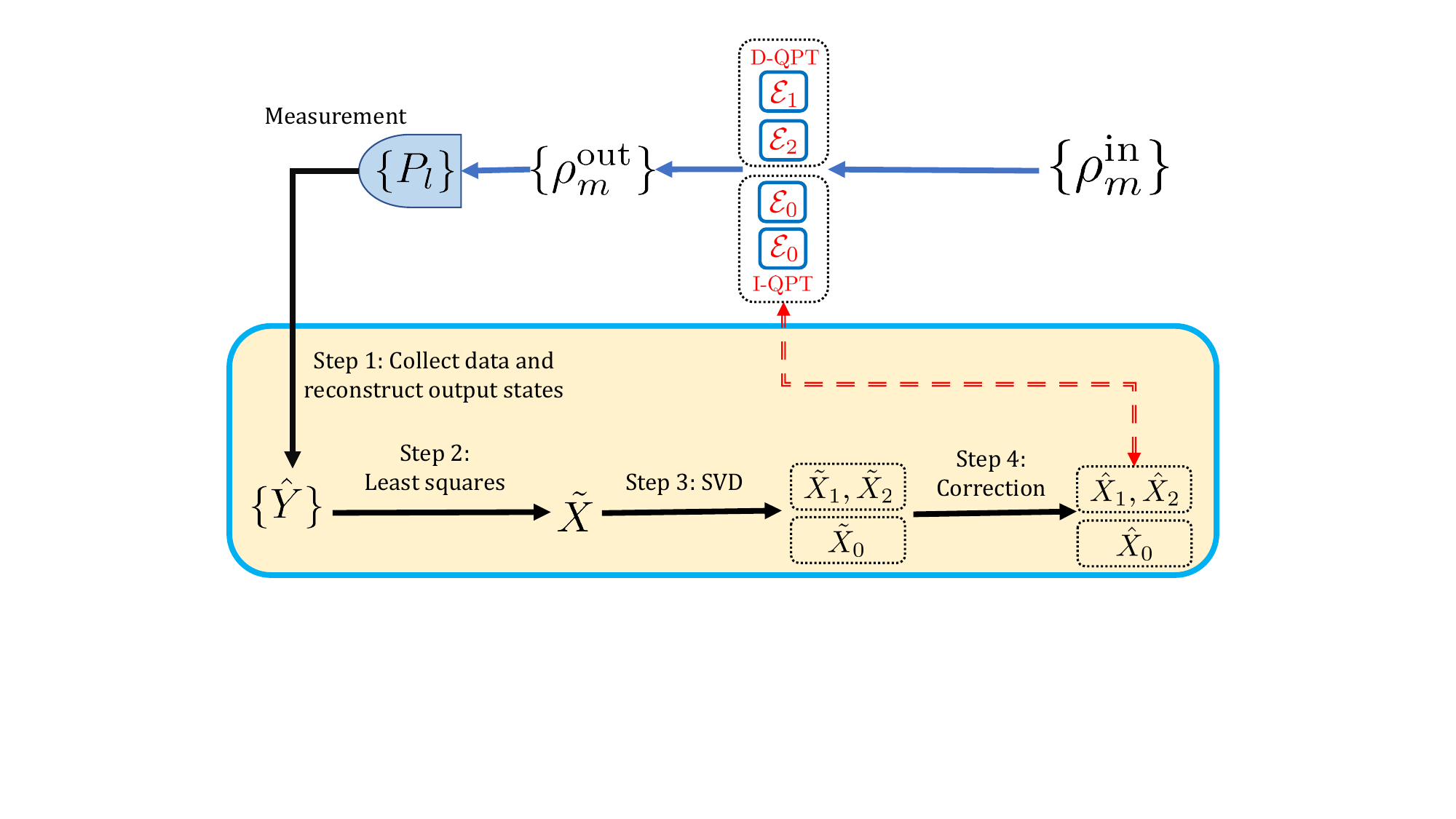}
	\centering{\caption{Steps in our closed-form algorithm for the generalized collective QPT.}\label{fpt}}
\end{figure}

Overall,
the procedures of our closed-form algorithm have four steps as outlined in Fig. \ref{fpt}.  Similar to \cite{xiao2024}, we can derive
the overall computational complexity $O(d_1^2d_2^2(ML+d_1^2 d_2^2 \min(d_1^4, d_2^4))$  for D-QPT and $O(d^4(ML+ d^8))$ for I-QPT, which is dominated by reconstructing theoutput states in Step 1, the least squares method in Step 2 and the SVD in Step 3 in Fig. \ref{fpt} for generalized collective QPT.

\subsection{Error analysis}

Throughout the paper, we use $N$ to denote the number of input state copies in each generalized collective protocol. In D/I-QST, $N$ corresponds to the number of state pairs. We also use $\mathbb E(\cdot)$ to denote the expectation with respect to all the possible measurement results.
Here we present the following theorems to describe the analytical MSE scaling using our closed-form algorithm for all generalized collective tomography procedures. First, we propose the following theorem for generalized collective QST.
\begin{theorem}\label{theorem1}
In the weakly informational-complete scenario,	 the MSE scalings of our algorithm satisfy $\mathbb E||\hat \rho_1-\rho_1||^2=O\left({1}/{N}\right)$ and $\mathbb E ||\hat \rho_2-\rho_2||^2=O\left({1}/{N}\right)$ for D-QST, and  $\mathbb E||\hat \rho_0-\rho_0||^2=O\left({1}/{N}\right)$ for I-QST.
\end{theorem}
\begin{IEEEproof}
	\subsubsection{Error in Steps 1 and 2}
	We collect data and utilize the least squares method. Thus for D-QST, we have
	\begin{equation}
		\begin{aligned}
&\mathbb E\left\|\tilde {x}-{\theta}_1 \otimes {\theta}_{2}\right\|^2\\
=&	\frac{1}{N}\operatorname{Tr}\Big[\left(\Phi^T \Phi\right)^{-1}\Phi^T \bar{C} \Phi\big(\Phi^T \Phi\big)^{-1}\Big]\\
= & O\left(\frac{1}{N}\right),
\end{aligned}
		\end{equation}
where $\bar{C}$ is a constant matrix determined by the true measurement result $Y$ \cite{Qi2013}. 
If we implement \eqref{x1}, we also have
\begin{equation}
		\begin{aligned}
	\mathbb E\left\|\tilde {x}-{\theta}_1 \otimes {\theta}_{2}\right\|^2&= 	\mathbb E\left\|\tilde {x}_b-\left({\theta}_1 \otimes {\theta}_{2} \right)_{2:d_1^2 d_2^2}\right\|^2 = O\left(\frac{1}{N}\right).\\
\end{aligned}
\end{equation}
Similarly for I-QST, we have
\begin{equation}
	\mathbb E\left\|\tilde {x}-{\theta}_0 \otimes {\theta}_{0}\right\|^2=O\left(\frac{1}{N}\right).
\end{equation}

\subsubsection{Error in Step 3}
In D-QST, we have
\begin{equation}\label{s31}
	\begin{aligned}
\mathbb E\left\|\tilde{\rho}-\rho_1 \otimes \rho_2\right\|^2=	\mathbb E\left\|\tilde {x}-{\theta}_1 \otimes {\theta}_{2}\right\|^2=O\left(\frac{1}{N}\right).
	\end{aligned}
\end{equation}
Since $\tilde \rho_1$ and $\tilde \rho_2$ minimize $\|\tilde \rho-\tilde\rho_1 \otimes \tilde \rho_2\|$ and using \eqref{s31}, we can obtain
\begin{equation}\label{cqst1}
	\mathbb E	\left\|\tilde{\rho}-\tilde\rho_1 \otimes \tilde \rho_2\right\|^2 \leq \mathbb E\left\|\tilde{\rho}- \rho_1 \otimes  \rho_2\right\|^2 =O\left(\frac{1}{N}\right).
\end{equation}
	Since 
\begin{equation}
	\begin{aligned}
		&\left\|\tilde \rho_1 \otimes \tilde \rho_2- \rho_1 \otimes  \rho_2\right\|\\
		\leq& \left\|\tilde{\rho}-\tilde\rho_1 \otimes \tilde \rho_2\right\|+\left\|\tilde{\rho}- \rho_1 \otimes  \rho_2\right\|,
	\end{aligned}
\end{equation}
and using \eqref{s31} and \eqref{cqst1},
we have 
\begin{equation}\label{rt1}
	\mathbb{E} \left\|\tilde \rho_1 \otimes \tilde \rho_2- \rho_1 \otimes  \rho_2\right\|^2=O\left(\frac{1}{N}\right).
\end{equation}
Similarly, in I-QST, we have
\begin{equation}\label{r2}
	\mathbb{E}\left\|\tilde \rho-\rho_0 \otimes \rho_0\right\|^2=O\left(\frac{1}{N}\right),
\end{equation}
and
\begin{equation}
	\mathbb{E}\left\|\tilde \rho_1 \otimes \tilde \rho_2-\rho_0 \otimes \rho_0\right\|^2=O\left(\frac{1}{N}\right).
\end{equation}

\subsubsection{Error in Step 4}
In D-QST,
given that $\sum_{i=1}^{d_1} \left(\rho_1\right)_{ii}=1$ and $\sum_{i=1}^{d_2} \left(\tilde \rho_1\right)_{ii}=1$, we obtain
\begin{equation}\label{rt2}
	\begin{aligned}
		&\left\|\tilde \rho_1 \otimes \tilde\rho_2- \rho_1 \otimes  \rho_2\right\|\\
		\geq &  \left\|\operatorname{diag}(\tilde\rho_1) \otimes\tilde \rho_2- \operatorname{diag}(\rho_1) \otimes  \rho_2\right\|\\
		=& \sqrt{\sum_{i=1}^{d_1} \left\|(\tilde{\rho}_1)_{ii} \tilde \rho_2 - ({\rho}_1)_{ii}  \rho_2 \right\|^2}\\
		\geq & \big\| \sum_{i=1}^{d_1} \left((\tilde{\rho}_1)_{ii} \tilde \rho_2 - ({\rho}_1)_{ii}  \rho_2 \right) \big\|/\sqrt{d_1}\\
		=& \|\tilde{\rho}_2-\rho_2\|/\sqrt{d_1},
	\end{aligned}
\end{equation}
where the first inequality comes from
\begin{equation}
		\begin{aligned}
	&\left\|\tilde \rho_1 \otimes \tilde\rho_2- \rho_1 \otimes  \rho_2\right\|^2=\left\|\operatorname{diag}(\tilde\rho_1) \otimes\tilde \rho_2- \operatorname{diag}(\rho_1) \otimes  \rho_2\right\|^2\\
	+ & \left\|(\tilde\rho_1-\operatorname{diag}(\tilde\rho_1)) \otimes\tilde \rho_2- (\rho_1-\operatorname{diag}(\rho_1)) \otimes  \rho_2\right\|^2,
		\end{aligned}
\end{equation}
and the second inequality comes from the Cauchy–Schwarz inequality.
Therefore, using \eqref{rt1} and \eqref{rt2}, we have
	\begin{equation}\label{rr1}
		\begin{aligned}
				\mathbb{E}\|\tilde{\rho}_2-\rho_2\|^2\leq d_1\mathbb{E} \left\|\tilde\rho_1 \otimes \tilde \rho_2- \rho_1 \otimes  \rho_2\right\|^2=O\left(\frac{1}{N}\right),
		\end{aligned}
	\end{equation}
	and thus 
	\begin{equation}\label{rho21}
	\mathbb{E}\|\tilde{\rho}_2^{\dagger}-\rho_2\|^2=O\left(\frac{1}{N}\right), \mathbb{E}\|(\tilde{\rho}_2)_{11}-(\rho_2)_{11}\|^2=O\left(\frac{1}{N}\right).
	\end{equation}
	Using \eqref{rt1}, we obtain
	\begin{equation}\label{rho22}
		\begin{aligned}
	\mathbb{E} \left\|(\tilde\rho_2)_{11}\tilde \rho_1  - (\rho_2)_{11} \rho_1 \right\|^2\!\leq \!	\mathbb{E} \left\|\tilde \rho_1 \otimes \tilde \rho_2- \rho_1 \otimes  \rho_2\right\|^2\!=\! O\left(\frac{1}{N}\right).
			\end{aligned}
	\end{equation}
	Without loss of generality, we assume $(\rho_2)_{11}>0$.
Since
	\begin{equation}\label{rho23}
		\begin{aligned}
				&|(\rho_{2})_{11}| \|\tilde\rho_1 -\rho_1\|\\
			=&\|(\rho_{2})_{11}\tilde\rho_1-  (\tilde\rho_{2})_{11}\tilde\rho_1+(\tilde\rho_{2})_{11}\tilde\rho_1-(\rho_{2})_{11}\rho_1\|\\
			\leq &\|\tilde\rho_1\| | (\tilde\rho_{2})_{11}-(\rho_{2})_{11}|+ \|(\tilde\rho_{2})_{11}\tilde\rho_1-(\rho_{2})_{11}\rho_1\|,
		\end{aligned}
	\end{equation}	
and using \eqref{rho21}, \eqref{rho22} and \eqref{rho23}, we have
\begin{equation}\label{aa2}
	\mathbb{E} \|\tilde\rho_1 -\rho_1\|^2=O\left(\frac{1}{N}\right),
\end{equation}
and thus $\mathbb{E} \|\tilde\rho_1^{\dagger} -\rho_1\|^2=O\left({1}/{N}\right)$. Therefore, we have
\begin{equation}
		\mathbb{E} \|\bar\rho_1 -\rho_1\|^2 =\mathbb{E} \left\| \frac{{\tilde \rho_1 +\tilde \rho_1^{\dagger}}}{2}-\rho_1\right\|^2 =O\left(\frac{1}{N}\right).
\end{equation}

Using Lemma \ref{lemma1} in Appendix \ref{appendixa} and \eqref{rr1}-\eqref{rho21}, we have
\begin{equation}\label{s32}
	\mathbb{E}|\operatorname{Tr}(\tilde{\rho}_2) -1|^2 =O\left(\frac{1}{N}\right).
\end{equation}
Let $\delta\triangleq\operatorname{Tr}(\tilde{\rho}_2) -1 $, and since
\begin{equation}
	\begin{aligned}
			\left\|\frac{\tilde{\rho}_2}{\operatorname{Tr}(\tilde{\rho}_2)}- \rho_2\right\|&=\left\|\frac{\tilde{\rho}_2- \rho_2(1+\delta)}{\operatorname{Tr}(\tilde{\rho}_2)}\right\|\\
			&\leq \left\|\frac{\tilde{\rho}_2- \rho_2}{\operatorname{Tr}(\tilde{\rho}_2)}\right\|+\left\|\frac{\rho_2\delta}{\operatorname{Tr}(\tilde{\rho}_2)}\right\|,
		\end{aligned}
\end{equation}
and using \eqref{rr1}, \eqref{s32},
we have
\begin{equation}\label{bar}
	\mathbb{E}\left\|\frac{\tilde{\rho}_2}{\operatorname{Tr}(\tilde{\rho}_2)}- \rho_2\right\|^2=O\left(\frac{1}{N}\right).
\end{equation}
and similarly
\begin{equation}\label{bar1}
	\mathbb{E}\left\|\frac{\tilde{\rho}_2^{\dagger}}{\operatorname{Tr}(\tilde{\rho}_2^{\dagger})}- \rho_2\right\|^2=O\left(\frac{1}{N}\right).
\end{equation}
Since
\begin{equation}
	\begin{aligned}
	\|\bar{\rho}_2-\rho_2\|&=\left\|\frac{1}{2}\left(\frac{{\tilde \rho_2 }}{\operatorname{Tr}({\tilde \rho_2})}+\frac{{\tilde \rho_2^{\dagger} }}{\operatorname{Tr}({\tilde \rho_2^{\dagger}})}\right)-\rho_2\right\|\\
	&\leq\frac{1}{2}\left\|\frac{\tilde{\rho}_2}{\operatorname{Tr}(\tilde{\rho}_2)}- \rho_2\right\|+\frac{1}{2}\left\|\frac{\tilde{\rho}_2^{\dagger}}{\operatorname{Tr}(\tilde{\rho}_2^{\dagger})}- \rho_2\right\|
		\end{aligned}
\end{equation}
and using \eqref{bar} and \eqref{bar1}, we have
\begin{equation}
	\mathbb{E} \|\bar\rho_2-\rho_2\|^2=O\left(\frac{1}{N}\right).
\end{equation}
We then implement the correction algorithms in \cite{effqst} and thus $	\mathbb{E}\left\|\hat{\rho}_1-\bar{\rho}_1\right\|^2=O(1/N)$, $\mathbb{E}\left\|\hat{\rho}_2-\bar{\rho}_2\right\|^2=O(1/N)$. Using \eqref{bar}, we finally obtain
\begin{equation}\label{aa3}
	\mathbb{E}\left\|\hat{\rho}_1-{\rho}_1\right\|^2=O\left(\frac{1}{N}\right), \;\;	\mathbb{E}\left\|\hat{\rho}_2-{\rho}_2\right\|^2=O\left(\frac{1}{N}\right).
\end{equation}

In I-QST, \eqref{aa3} still holds. Since
\begin{equation}
	\|\rho_0 -\frac{\hat{\rho}_1+\hat{\rho}_2}{2}\| \leq \frac{1}{2}\|\hat{\rho}_1-\rho_0\|+\frac{1}{2}\|\hat{\rho}_2-\rho_0\|,
\end{equation}
we obtain
\begin{equation}\label{rhoe0}
		\mathbb{E}\|\hat{\rho}_0-\rho_0\|^2=O\left(\frac{1}{N}\right).
\end{equation}
\end{IEEEproof}

\begin{remark}
If we adopt the quantum relative entropy $D(\cdot \| \cdot)$ as in Remark~\ref{rem4} 
and apply Lemma~\ref{lemma2} in Appendix~\ref{appendixa}, the mean-square errors of the optimal solution satisfy 
$\mathbb{E}\| \tilde{\rho}_1 - \rho_1 \|^2 = O(1/N)$ 
and $\mathbb{E}\| \tilde{\rho}_2 - \rho_2 \|^2 = O(1/N)$. 
Furthermore, by Lemma~\ref{lemma5} in Appendix~\ref{appendixa}, we obtain 
$\mathbb{E}[D( \tilde{\rho}_1 \| \rho_1 )] = O(1/N)$ 
and $\mathbb{E}[D(\tilde{\rho}_2  \| \rho_2)] = O(1/N)$ 
when ${\rho}_1$ and ${\rho}_2$ have full rank. Otherwise, if ${\rho}_1$ and ${\rho}_2$  are rank-deficient, the quantum relative entropy may be infinite \cite{wilde2013quantum}.
\end{remark}

We then propose the following theorem for generalized collective QDT.
\begin{theorem}\label{theorem2}
	In the weakly informational-complete scenario,	 the MSE scalings of our algorithm satisfy $\mathbb E \sum_{l=1}^{L}||\hat P_l-P_l||^2=O\left({1}/{N}\right)$ and $\mathbb E \sum_{k=1}^{K}||\hat Q_k-Q_k||^2=O\left({1}/{N}\right)$ for D-QDT and $\mathbb E \sum_{l=1}^{L}||\hat P_l-P_l||^2=O\left({1}/{N}\right)$ for I-QDT.
\end{theorem}
\begin{IEEEproof}
For D-QDT, similar to collective QST, we have
\begin{equation}\label{d1}
\mathbb{E}\|\tilde R_{lk} - P_{l} \otimes Q_{k}\|^2
\leq \frac{M}{4N} \operatorname{Tr}\!\left[\left(\Theta^T \Theta\right)^{-1}\right]
= O\!\left(\frac{1}{N}\right).
\end{equation}  
Similar to \eqref{rt1}, it follows that
\begin{equation}\label{q20}
\mathbb{E}\|\tilde P_{l} \otimes \tilde Q_{k} - P_{l} \otimes Q_{k}\|^2 = O\!\left(\frac{1}{N}\right),
\end{equation}
and from \eqref{d1}
\begin{equation}\label{q21}
\mathbb{E}\|\tilde R_{lk} - R_{lk}\|^2
= \mathbb{E}\|\tilde R_{lk} - P_{l} \otimes Q_k\|^2
= O\!\left(\frac{1}{N}\right).
\end{equation}
Using \eqref{q20}, we obtain
\begin{equation}\label{qt1}
\mathbb{E}\|(\tilde{P}_l)_{ii}\tilde{Q}_k - ({P}_l)_{ii}{Q}_k\|^2
= O\!\left(\frac{1}{N}\right).
\end{equation}
Since 
\[
\sum_{l=1}^{L} \operatorname{Tr}(\tilde P_l) = \sum_{l=1}^{L} \operatorname{Tr}(P_l) = d,
\]
and
\begin{equation}
\left\|\sum_{i=1}^{d_1} \big((\tilde P_l)_{ii}\tilde Q_k - (P_l)_{ii}Q_k\big)\right\|
\leq \sum_{i=1}^{d_1} \|(\tilde P_l)_{ii}\tilde Q_k - (P_l)_{ii}Q_k\|,
\end{equation}
it follows from \eqref{qt1} that
\begin{equation}\label{q3}
\mathbb{E}\|\tilde Q_k - Q_k\|^2 = O\!\left(\tfrac{1}{N}\right).
\end{equation}
Consequently, we also have 
\begin{equation}
\mathbb{E}\|\tilde Q_k^{\dagger} - Q_k\|^2 = O\!\left(\tfrac{1}{N}\right).    
\end{equation}
 Hence,
\begin{equation}\label{81}
\mathbb{E}\|\bar{Q}_k - Q_k\|^2
= \mathbb{E}\left\|\tfrac{\tilde{Q}_k + \tilde{Q}_k^{\dagger}}{2} - Q_k\right\|^2
= O\!\left(\frac{1}{N}\right).
\end{equation}
Similarly, we can show $\mathbb{E}\|\bar{P}_l - P_l\|^2 = O(1/N)$.  

By applying the Stage-2 correction algorithm in \cite{wang2019twostage}, which preserves the MSE scaling, we obtain
\begin{equation}\label{q4}
\mathbb{E}\|\hat{P}_l - \bar{P}_l\|^2 = O\!\left(\frac{1}{N}\right),
\mathbb{E}\|\hat{Q}_k - \bar{Q}_k\|^2 = O\!\left(\frac{1}{N}\right).
\end{equation}
Combining \eqref{81} and \eqref{q4} yields
\begin{equation}
\mathbb{E}\|\hat{P}_l - P_l\|^2 = O\!\left(\frac{1}{N}\right), 
\mathbb{E}\|\hat{Q}_k - Q_k\|^2 = O\!\left(\frac{1}{N}\right).
\end{equation}
Therefore,
\begin{equation}
\mathbb{E}\sum_{l=1}^{L}\|\hat{P}_l - P_l\|^2 = O\!\left(\frac{1}{N}\right), \;
\mathbb{E}\sum_{k=1}^{K}\|\hat Q_k - Q_k\|^2 = O\!\left(\frac{1}{N}\right).
\end{equation}

For I-QDT, by an argument analogous to the proof for D-QDT, we similarly obtain
\begin{equation}
\mathbb{E}\sum_{l=1}^{L}\|\hat{P}_l - P_l\|^2 = O\!\left(\frac{1}{N}\right).
\end{equation}
\end{IEEEproof}

\begin{theorem}\label{theorem3}
	In the weakly informational-complete scenario,	 the MSE scalings of our algorithm satisfy $\mathbb E||\hat X_1-X_1||^2=O\left({1}/{N}\right)$ and $\mathbb E ||\hat X_2-X_2||^2=O\left({1}/{N}\right)$ for D-QPT and  $\mathbb E||\hat X_0-X_0||^2=O\left({1}/{N}\right)$ for I-QPT.
\end{theorem}

We omit the proof here because the proofs for D-QPT and I-QPT are similar to those of Theorems \ref{theorem1} and \ref{theorem2}, respectively.

\subsection{Extension to $n$-copy $(n\geq 3)$ cases}
It is straightforward to extend our closed-form algorithm to $n$-copy $(n\geq 3)$ scenarios. For $n$-copy $(n\geq 3)$ collective QST, a similar procedure can be applied. In D-QST, we first use the least squares method, and then solve the optimization problem $\min_{\tilde{\rho}_1, \cdots, \tilde{\rho}_n} \|\tilde{\rho} - \tilde{\rho}_1 \otimes \cdots \otimes \tilde{\rho}_n\|$. We start by solving $\min_{\tilde{\rho}_1, \tilde{\rho}_{2:n}} \|\tilde{\rho} - \tilde{\rho}_1 \otimes \tilde{\rho}_{2:n}\|$, where $\tilde{\rho}_{2:n} \in \mathbb{C}^{(n-1)d \times (n-1)d}$, and then proceed with $\min_{\tilde{\rho}_2, \tilde{\rho}_{3:n}} \|\tilde{\rho}_{2:n} - \tilde{\rho}_2 \otimes \tilde{\rho}_{3:n}\|$. This step-by-step approach allows us to obtain a closed-form solution using the SVD like Problem \ref{subproblem2}. For error analysis, similar to \eqref{rr1}, we have
\begin{equation}
	\mathbb{E}\|\tilde{\rho}_1 - \rho_1\|^2 = O\left(\frac{1}{N}\right), \; \mathbb{E}\|\tilde{\rho}_{2:n} - \rho_{2:n}\|^2 = O\left(\frac{1}{N}\right).
\end{equation}
Through step-by-step decoupling of the tensor product, we can further show
\begin{equation}
	\mathbb{E}\|\tilde{\rho}_j - \rho_j\|^2 = O\left(\frac{1}{N}\right), \; 1 \leq j \leq n.
\end{equation}
We then implement the correction algorithms from \cite{effqst} and thus $	\mathbb{E}\left\|\hat{\rho}_j-{\rho}_j\right\|^2=O(1/N), 1 \leq j \leq n$.
For I-QST, we compute $\hat{\rho}_0 = \frac{\sum_{j=1}^{n} \hat{\rho}_j}{n}$, and similar to \eqref{rhoe0}, we obtain
\begin{equation}
	\mathbb{E}\|\hat{\rho}_0 - \rho_0\|^2 = O\left(\frac{1}{N}\right).
\end{equation}

For $n$-copy $(n\geq 3)$ generalized collective QDT and QPT, we also begin with the least squares method. In D-QDT and D-QPT, we can similarly achieve  closed-form solutions using step-by-step SVD decomposition. As with D-QST, we can prove that the MSE scalings are $O(1/N)$. For I-QDT and I-QPT, one  solution is to first use step-by-step SVD decomposition similar to D-QDT and D-QPT for $n\geq3$, followed by averaging all the estimates. Similar to I-QST, the MSE still scales as $O(1/N)$.

\section{Sum of squares optimization}\label{sosop}
The above closed-form solution is computationally efficient, but it may be suboptimal due to the relaxation introduced by decomposing the problem into two subproblems. To pursue an optimal solution that respects the original tensor structure in collective tomography, we propose a sum of squares (SOS) optimization approach in this section.

Testing the polynomial nonnegativity condition
\begin{equation}
	p(x):=p(x_1,\ldots,x_n) \geq 0
\end{equation}
for all $x\in \mathbb{R}^{n}$ is NP-hard, even when $p(x)$ is of degree $4$. However, a more tractable sufficient condition for the nonnegativity of  $p(x)$ is for it to be a sum of squares (SOS) polynomial. This means that $p(x)$  can be expressed as
\begin{equation}
	p(x)=\sum_{i=1}^r f_i^2(x)
\end{equation}
for some polynomials $ f_i(x)$.
Determining whether a polynomial is a sum of squares reduces to solving a semidefinite program (SDP), a class of convex optimization problems for which efficient numerical solution methods are available.
Since the cost function of the collective tomography problem is a polynomial that is always non-negative, we can use SOS optimization to solve this problem. The problem of obtaining a lower bound for the global minimum of a polynomial function using SOS optimization is discussed in detail in \cite{parrilo2001}.

For collective QST, we need to consider the constraints on quantum states. The Hermitian and the unit trace constraints on quantum states have been effectively satisfied by selecting the basis  $\{\Omega_{i}\}_{i=1}^{d_1^{2}}$ and $\{\Xi_{j}\}_{j=1}^{d_2^{2}}$. 
The positive semidefinite constraints of quantum states  are complicated and can be characterized by a semi-algebraic set. These constraints have been extensively addressed in literature \cite{Kimura2003,1440563}, and we can use the following lemma to describe the physical set characterizing the parameterized vector of a quantum state, e.g., $\theta_0$.
\begin{lemma} (\cite{Kimura2003,1440563})
	Define $k_p^{d}(\rho)$, $p=2,\cdots,d$  recursively by
	\begin{equation}
		pk_p^{d}(\rho)=\sum_{f=1}^p(-1)^{f-1}\operatorname{Tr}(\rho^f)k_{p-f}^{d}(\rho)
	\end{equation}
	with $k_0^{d}=k_1^{d}=1$. Define the semi-algebraic set 
	\begin{equation}
		\mathcal{K}_{d}=\{\theta_0\in\mathbb{R}^{d^2}:k_p^{d}(h(\theta_0))\geq0, p=2,\cdots, d\}.
	\end{equation}
	Then $h(\cdot)$ (defined after \eqref{rho1}) is an isomorphism between $\mathcal{K}_{d}$ and $\mathcal{S}_{d}$.
\end{lemma}

Thus, we propose to solve Problem \ref{p1} by addressing the following optimization problem:
\begin{equation}\label{sosqs}
	\begin{aligned}
		\min&\;(-\gamma)\\
		\text{s.t. } &||\hat Y-\Phi\left(\theta_1\otimes \theta_2\right)||^2-\gamma \text{ is SOS,}\\
		& (\theta_1)_1=\frac{1}{\sqrt{d_1}}, (\theta_2)_1=\frac{1}{\sqrt{d_2}}, {\theta_1}\in \mathcal{K}_{d_1},   {\theta_2}\in \mathcal{K}_{d_2},
	\end{aligned}
\end{equation}
and  similarly, Problem \ref{p2} can be formulated as:
\begin{equation}\label{sosqs2}
	\begin{aligned}
		\min&\;(-\gamma)\\
		\text{s.t. } &||\hat Y-\Phi\left(\theta_0\otimes \theta_0\right)||^2-\gamma \text{ is SOS,}\\
		& (\theta_0)_1=\frac{1}{\sqrt{d}}, {\theta_0}\in \mathcal{K}_{d}.   
	\end{aligned}
\end{equation}

For collective QDT, the Hermitian constraints on $\{P_l\}_{l=1}^{L}$ and $\{Q_k\}_{k=1}^{K}$ have also been effectively satisfied by selecting the bases $\{\Omega_{i}\}_{i=1}^{d_1^{2}}$ and $\{\Xi_{j}\}_{j=1}^{d_2^{2}}$. Furthermore, the completeness constraints on $\{P_l\}_{l=1}^{M}$ and $\{Q_k\}_{k=1}^{K}$ can be expressed as
\begin{equation}
	\begin{aligned}
			\sum_{l=1}^{L} P_l&=I_{d_1} \Leftrightarrow	\sum_{l=1}^L \phi_l=[\sqrt d_1,0,\cdots, 0],\\
			\sum_{k=1}^{K} Q_k&=I_{d_2} \Leftrightarrow	\sum_{k=1}^K \varphi_k=[\sqrt d_2,0,\cdots, 0].
	\end{aligned}
\end{equation}
For each POVM element $P_l$, we can also normalize it to a density matrix and obtain the similar semi-algebraic set. Hence, $P_l$ is positive semidefinite if and only if $$\frac{\phi_l}{\sqrt{d_1}\phi_{l}^{1}} \in \mathcal{K}_{d}.$$
Therefore, we can formulate generalized collective QDT in Problem \ref{pd1} as
\begin{equation}\label{sosqd}
	\begin{aligned}
		\min&\;(-\gamma)\\
		\text{s.t. } &\sum_{l.k=1}^{L,K}||\hat Y_{lk}-\Phi\left(\phi_l\otimes \varphi_k\right)||^2-\gamma \text{ is SOS,}\\
	&\sum_{l=1}^L \phi_l=[\sqrt d_1,0,\cdots, 0], \sum_{k=1}^K \varphi_k=[\sqrt d_2,0,\cdots, 0],\\
		& \frac{\phi_l}{\sqrt{d_1}\phi_{l}^{1}}\in \mathcal{K}_{d_1}, \frac{\varphi_k}{\sqrt{d_2}\varphi_{k}^{1}} \in \mathcal{K}_{d_2} \text{ for } 1\leq l \leq L, 1\leq k \leq K,
	\end{aligned}
\end{equation}
and  Problem \ref{pd2} as
\begin{equation}\label{sosqd2}
	\begin{aligned}
		\min&\;(-\gamma)\\
		\text{s.t. } &\sum_{l,k=1}^{L,L}||\hat Y_{lk}-\Phi\left(\phi_l\otimes \phi_k\right)||^2-\gamma \text{ is SOS,}\\
		&\sum_{l=1}^L \phi_l=[\sqrt d,0,\cdots, 0], \\
		& \frac{\phi_l}{\sqrt{d}\phi_{l}^{1}}  \in \mathcal{K}_{d} \text{ for } 1\leq l \leq L.
	\end{aligned}
\end{equation}

Similarly, for collective QPT, we can also formulate it as 
\begin{equation}\label{sosqp}
	\begin{aligned}
		\min&\;(-\gamma)\\
		\text{s.t. } &\big\| \hat Y-\mathcal{B}(\operatorname{vec}(X_1)\otimes \operatorname{vec}(X_2))\big\|^2-\gamma \text{ is SOS,}\\
		&\operatorname{Tr}_{1} (X_1)\leq I_{d_1}, \operatorname{Tr}_{1} (X_2)\leq I_{d_2}\\
		& \frac{X_1}{\operatorname{Tr}(X_1)} \in \mathcal{K}_{d_1^2}, \frac{X_2}{\operatorname{Tr}(X_2)} \in \mathcal{K}_{d_2^2},
	\end{aligned}
\end{equation}
and for TP processes, the partial trace constraint becomes $\operatorname{Tr}_{1} (X_1)= I_{d_1}, \operatorname{Tr}_{1} (X_2)= I_{d_2}$.
For I-QPT, we just need to change $X_1$ and $X_2$ to $X_0$.

All the above constrained polynomial optimization problems can be efficiently solved using \texttt{findbound} function in \texttt{SOSTOOLS}~\cite{sostools}. This function yields a lower bound~$\gamma$ for the cost function. When \texttt{findbound} also returns values for the optimization variables (e.g.,~$\theta_0$), it implies that the lower bound is attainable, and hence~$\gamma$ corresponds to the global minimum of the cost function. However, in some cases, \texttt{findbound} may fail to return feasible variable values if the lower bound is unattainable. In such cases, the obtained lower bound can still serve as a reference for evaluating candidate estimates, allowing us to select the one closest to~$\gamma$.
 Different regularization matrices, as outlined in \eqref{re}, can be used to obtain possible candidates.
Nevertheless, when \texttt{findbound} function does return values for the optimization variables, it indicates that these values achieve the lower bound. Consequently, this lower bound represents the minimum value of the cost function in the original problem.

Based on the effectiveness of SOS optimization, we propose using SOS tests with the property that, even in the weakly informational-incomplete scenario, SOS can also provide a unique optimal solution. For collective QST, after preparing the measurement operators, we can directly implement them in experiments and use the experimental data to solve the optimization problems given by \eqref{sosqs} or \eqref{sosqs2}. If we obtain a solution, it is guaranteed to be optimal. Otherwise, additional measurement operators may be needed to determine a unique solution. As demonstrated in \cite{Hou2018,zhou2023experimental}, using two-copy \eqref{21}-\eqref{22} and three-copy \eqref{31} collective measurement can yield a unique optimal solution with $L=5, 9$ different measurement operators. An interesting open problem is how to choose further measurement operators when a unique optimal solution cannot be found through SOS optimization. Addressing this challenge could lead to more effective strategies for selecting measurement operators in collective QST. A similar SOS test can also be applied to collective QDT and collective QPT, which may require fewer probe states and input states, therefore being more practically feasible in experimental settings.

\section{Examples and numerical results}\label{example}
In this section, we present illustrative examples and numerical simulations to demonstrate the implementation and performance of our proposed generalized collective tomography.
\subsection{Illustrative examples}\label{secexample}
In this subsection, we consider several illustrative examples to demonstrate the implementation of our generalized collective tomography method. We begin by employing prior information indicating that the unknown state is pure for QST, a useful property in quantum technologies, simplifying the constraints in SOS optimization. Similarly, we consider that the POVM elements are projective in QDT.
Finally, we consider the problems of collective Hamiltonian/unitary tomography.

\subsubsection{Pure state  and projective measurement tomography}
 Pure states are important quantum resources and widely applied in experiments. Thus, we may assume prior knowledge that the unknown states $\rho_i=|\psi_i\rangle\langle\psi_i|\; (i=1,2)$ are pure states.  
  Using \eqref{property2}, we have
 \begin{equation}\label{trho}
 	\begin{aligned}
 	\operatorname{vec}(|\psi_i\rangle\langle\psi_i|)=|\psi_i\rangle^{*}\otimes |\psi_i\rangle, \;i=1,2.
 	 \end{aligned}
 \end{equation}
 Using \eqref{trho} and Lemma \ref{pro1} in Appendix \ref{appendixa}, we have
 \begin{equation}
 	\begin{aligned}
 		&\operatorname{vec}(\rho_1\otimes \rho_2 )=\left(I_{d_1} \otimes K_{d_2d_1} \otimes I_{d_2}\right)\!(\operatorname{vec} (\rho_1) \otimes \operatorname{vec} (\rho_2))\\
 		=&\left(I_{d_1} \otimes K_{d_2d_1} \otimes I_{d_2}\right)\!(|\psi_1\rangle^{*}\otimes |\psi_1\rangle \otimes |\psi_2\rangle^{*}\otimes |\psi_2\rangle).
 	\end{aligned}
 \end{equation}
 Therefore, we have
  \begin{equation}
 	\begin{aligned}
 	&\quad p_l=\operatorname{Tr}(P_l (\rho_1 \otimes \rho_2))\\
 	&= (\operatorname{vec}(P_l))^{\dagger} \operatorname{vec}(\rho_1 \otimes \rho_2) \\
 		&=(\operatorname{vec}(P_l))^{\dagger} \left(I_{d_1} \otimes K_{d_1d_2} \otimes I_{d_2}\right)(|\psi_1\rangle^{*}\otimes |\psi_1\rangle \otimes |\psi_2\rangle^{*}\otimes |\psi_2\rangle).
 	\end{aligned}
 \end{equation}
Denote $\upsilon_l =\left(I_{d_1} \otimes K_{d_2d_1} \otimes I_{d_2}\right)^{T} \operatorname{vec}(P_l)^{*}$ and 
$\Upsilon=[\upsilon_1, \upsilon_2, \cdots, $ $\upsilon_L]^{T}$.
Hence, Problem \ref{p1} becomes as follows:
\renewcommand{\theproblem}{\arabic{problem}}
\addtocounter{problem}{0}
\begin{problem}\label{problem31}
	Given the matrix $ \Upsilon$ and experimental data $\hat Y$, solve $$\min_{|\psi_1\rangle,|\psi_2\rangle } ||\hat Y- \Upsilon(|\psi_1\rangle^{*}\otimes |\psi_1\rangle \otimes |\psi_2\rangle^{*}\otimes |\psi_2\rangle)||^2$$ where $|\psi_1\rangle \in \mathbb{C}^{d_1}, |\psi_2\rangle \in \mathbb{C}^{d_2}$ are unit vectors.
\end{problem}

To solve this problem with a closed-form solution, we can first implement Algorithm \ref{A1} in Sec. \ref{sec31} and obtain $\hat\rho_1$ and $\hat\rho_2$. Assuming the spectral decomposition $\hat\rho_1=\hat U_1 \operatorname{diag}(\hat\lambda_1, \cdots, \hat\lambda_{d_1}) \hat U_1^{\dagger} $ where $\hat\lambda_1\geq \cdots \geq \hat\lambda_{d_1}$,  the final estimate of the input pure state is 
\begin{equation}\label{pure}
	\hat \rho_1^{\prime}=\hat U_1\operatorname{diag}(1, 0, \cdots, 0) \hat U_1.
\end{equation}
Similarly, we can obtain the final estimate $\hat \rho_2^{\prime}$ and prove that the final MSE scalings of the quantum states are still  $O(1/N)$.

Although the algorithm is still closed-form and can achieve $O(1/N)$ MSE scaling, it does not utilize the pure state prior information and only satisfies the pure constraint by correction. Alternatively, we can also formulate it as an SOS problem 
\begin{equation}\label{sospure}
	\begin{aligned}
		\min&\;(-\gamma)\\
		\text{s.t. } &||\hat Y- \Upsilon(|\psi_1\rangle^{*}\otimes |\psi_1\rangle \otimes |\psi_2\rangle^{*}\otimes |\psi_2\rangle)||^2-\gamma \text{ is SOS,}\\
		& |\psi_1\rangle \in \mathbb{C}^{d_1}, |\psi_2\rangle \in \mathbb{C}^{d_2}, \||\psi_1\rangle\|=\||\psi_2\rangle\|=1,\\
	\end{aligned}
\end{equation}
which utilizes the pure state prior information in the cost function and thus does not need to consider the unit trace and positive semidefinite constraint anymore.

For collective QDT in Problem \ref{pd1},
if the POVM is projective, i.e., $P_l=|\phi_l\rangle\langle\phi_l|$, we can correct the eigenvalues of  $\{\hat P_l\}$ as \eqref{pure} and the MSE still scales as $O(1/N)$. We can also formulate it as an SOS optimization,
where the cost function becomes
	\begin{equation}
	\begin{aligned}
		&\sum_{l,k=1}^{L,K} ||\hat Y_{lk}-\Theta \big(|\phi_l\rangle^{*}\otimes |\phi_l\rangle\otimes |\varphi_k\rangle^{*}\otimes |\varphi_k\rangle\big)||^2-\gamma \text{ is SOS,}\\
		&  \||\phi_l\rangle\|=\||\varphi_k\rangle\|=1, \forall\; 1\leq l \leq L, 1 \leq k \leq K,\\
		& \sum_{l=1}^{L} |\phi_l\rangle \langle \phi_l|=I_{d_1}, \sum_{k=1}^{K} |\varphi_k\rangle \langle \varphi_k|=I_{d_2}.
	\end{aligned}
\end{equation}
Hence we do not need to consider the positive semidefinite constraint on POVM elements anymore.

\subsubsection{Unitary/Hamiltonian tomography}
If the unknown quantum process $\mathcal{E}_1$ is unitary driven by the Hamiltonian $H_1$, we have $X_1=\operatorname{vec}(G_1)\operatorname{vec}(G_1)^{\dagger}$, where $G_1$ is a unitary operator \cite{8022944}. Thus, using \eqref{property2}, we have
\begin{equation}
	\operatorname{vec}(X_1)=\operatorname{vec}(G_1)^{*} \otimes \operatorname{vec}(G_1).
\end{equation}
Since $\operatorname{rank}(X_1)=1$,  we can apply a similar procedure to that for the pure state case by correcting the eigenvalues of $X_1$.
After this correction, it is not difficult to prove that the MSE scaling $\mathbb{E}\|\hat G_1- G_1\|^2$ is still $O(1/N)$. Using the Hamiltonian reconstructing algorithm in \cite{8022944}, the MSE scaling of the Hamiltonian $\mathbb{E}\|\hat H_1- H_1\|^2$ is still $O(1/N)$.

Using the SOS optimization, for D-QPT, we have 
\begin{equation}\label{sosp1}
	\begin{aligned}
		\min&\;(-\gamma)\\
		\text{s.t. } &\big\| \hat Y-\mathcal{B}\big(\operatorname{vec}(G_1)^{*} \otimes \operatorname{vec}(G_1)\\&\otimes
		\operatorname{vec}(G_2)^{*} \otimes \operatorname{vec}(G_2)\big)\big\|^2-\gamma \text{ is SOS},\\
		& G_1 G_1^{\dagger}=I_{d_1}, G_2 G_2^{\dagger}=I_{d_2},
	\end{aligned}
\end{equation}
and, for I-QPT, we have
\begin{equation}\label{sosp2}
	\begin{aligned}
		\min&\;(-\gamma)\\
		\text{s.t. } &\big\| \hat Y-\mathcal{B}\big(\operatorname{vec}(G_0)^{*} \otimes \operatorname{vec}(G_0)\\&\otimes
		\operatorname{vec}(G_0)^{*} \otimes \operatorname{vec}(G_0)\big)\big\|^2-\gamma \text{ is SOS},\\
		& G_0 G_0^{\dagger}=I_{d}.
	\end{aligned}
\end{equation}
After obtaining $G_1$ and $G_2$, or $G_0$, we can also reconstruct  the Hamiltonian $H_1$ and $H_2$, or $H_0$, using the algorithm in \cite{8022944}.

\subsection{Numerical results}\label{sec6}
In this subsection, we present three numerical examples: two-copy and three-copy collective QST, two-copy collective QDT, and two-copy collective phase damping process tomography. The simulations were conducted on a laptop with an Intel i9-13980HX processor and 64 GB of DDR5 memory. For each data point in the figures shown in this section, we repeated our algorithm 100 times and computed the mean to obtain the MSE and error bars. The random unitary matrices and random probe states used in these simulations were generated using the algorithms in \cite{MISZCZAK2012118, qetlab}.
\subsubsection{Two-copy and three-copy collective QST}
\begin{figure}[!t]
	\centering
	\includegraphics[width=3.3in]{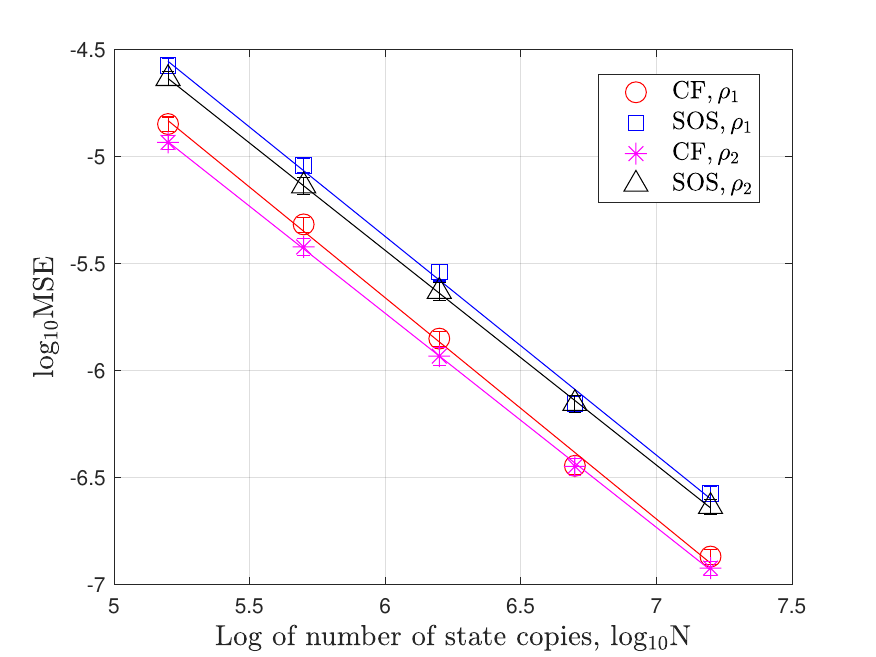}
	\centering{\caption{Log-log plot of MSE versus  the total number of state copies $N$  for D-QST using the closed-form (CF) solution  and SOS optimization with MUB measurements \eqref{mubbase}.}\label{f1}}
\end{figure}

\begin{figure}[!t]
	\centering
	\includegraphics[width=3.3in]{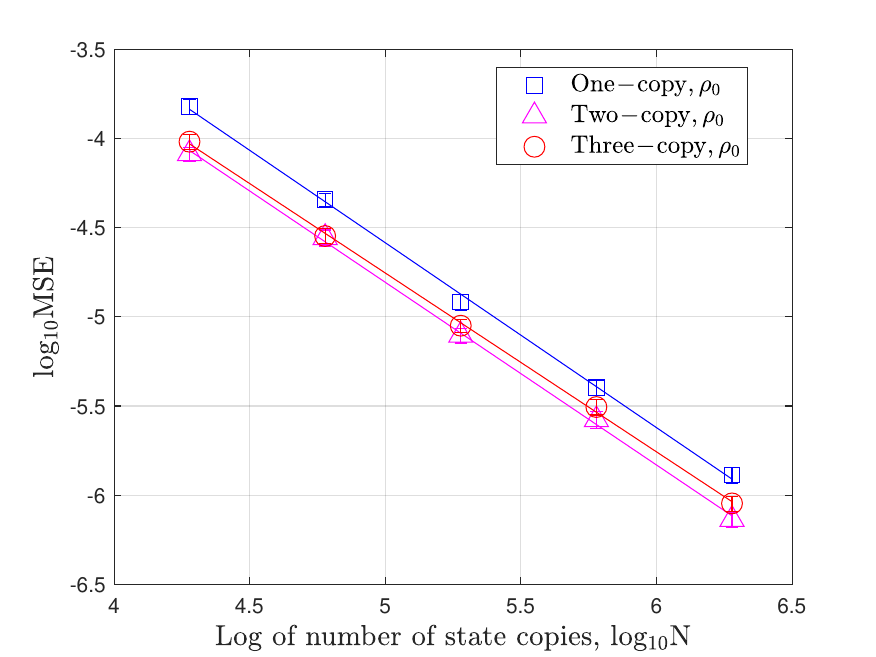}
	\centering{\caption{Log-log plot of MSE versus  the total number of state copies $N$  for individual SIC-POVM \eqref{sic}, two-copy collective POVM \eqref{21}-\eqref{22}  and three-copy collective POVM \eqref{31} using SOS optimization in I-QST.}\label{f2}}
\end{figure}

For D-QST,
let the unknown initial quantum states be
\begin{equation}
		\begin{aligned}
	\rho_1&=U_1\operatorname{diag}(0.9,0.1)U_1^\dagger, \\
	\rho_2&=U_2\operatorname{diag}(0.8,0.2)U_2^\dagger,
\end{aligned}
\end{equation}
where $ U_1 $ and $ U_2 $ are randomly generated unitary matrices  that are subsequently fixed throughout the process. For this one-qubit example of SOS optimization, $\theta_1, \theta_2 \in \mathcal{K}$ are equivalent to $\sum_{i=2}^{4} (\theta_1)_i^2 \leq \frac{1}{2},\; \sum_{i=2}^{4} (\theta_2)_i^2 \leq \frac{1}{2}$ because of the Bloch sphere representation~\cite{Kimura2003,1440563}.
Thus, the SOS optimization can be reformulated as
\begin{equation}\label{sosqsone}
	\begin{aligned}
		\min&\;(-\gamma)\\
		\text{s.t. } &||\hat Y-\Phi\left(\theta_1\otimes \theta_2\right)||^2-\gamma \text{ is SOS,}\\
		& (\theta_1)_1=(\theta_2)_1=\frac{1}{\sqrt{2}},\\
		& \sum_{i=2}^{4} (\theta_1)_i^2 \leq \frac{1}{2},\;\; \sum_{i=2}^{4} (\theta_2)_i^2 \leq \frac{1}{2}.
	\end{aligned}
\end{equation}
We utilize mutually unbiased basis (MUB) measurements as defined in \eqref{mubbase} for this case, and the results are presented in Fig. \ref{f1}. The MSE scalings of both the closed-form (CF) solution and the SOS optimization are $O(1/N)$,  which is consistent with Theorem \ref{theorem1}. Additionally, the MSEs of the CF solution are slightly smaller than those of the SOS optimization. This occurs because minimizing the cost function does not necessarily minimize the MSE of the parameters.

For I-QST, let the unknown state  be 
\begin{equation}
	\rho_0=U\operatorname{diag}(1,0)U^\dagger
\end{equation}
which is a pure state and $U$ is a randomly generated  matrix and then fixed in the simulation.
We then implement the individual symmetric, informationally complete POVM (SIC-POVM) \eqref{sic}, the two-copy collective POVM \eqref{21}-\eqref{22},  and the three-copy collective POVM \eqref{31}, as detailed in Appendix \ref{appb}. To ensure that the total number of resources is consistent across the different POVM implementations, we set $N_0$ a shared value among these different schemes while
$N=6N_0$ for individual SIC-POVM, $N=3N_0$ for the two-copy collective POVM and $N=2N_0$ for the three-copy collective POVM. In this simulation, {\ttfamily{findbound}}  can always output values of the optimization variable $\theta_0$ and thus the lower bound $\gamma$ is the minimum value of the cost function.


The results obtained via SOS optimization are shown in Fig.~\ref{f2}, where all MSE scalings follow $O(1/N)$ and the performance differences remain minor. The three-copy collective POVM exhibits MSE behavior comparable to that of the two-copy counterpart, indicating that in practical scenarios the two-copy implementation may already suffice, providing a simpler alternative while still achieving lower MSE. In contrast, the experimental results in Ref.~\cite{zhou2023experimental} demonstrate that the three-copy collective measurement can attain a higher fidelity.

\subsubsection{Two-copy collective QDT}
\begin{figure}[!t]
	\centering
	\includegraphics[width=3.3in]{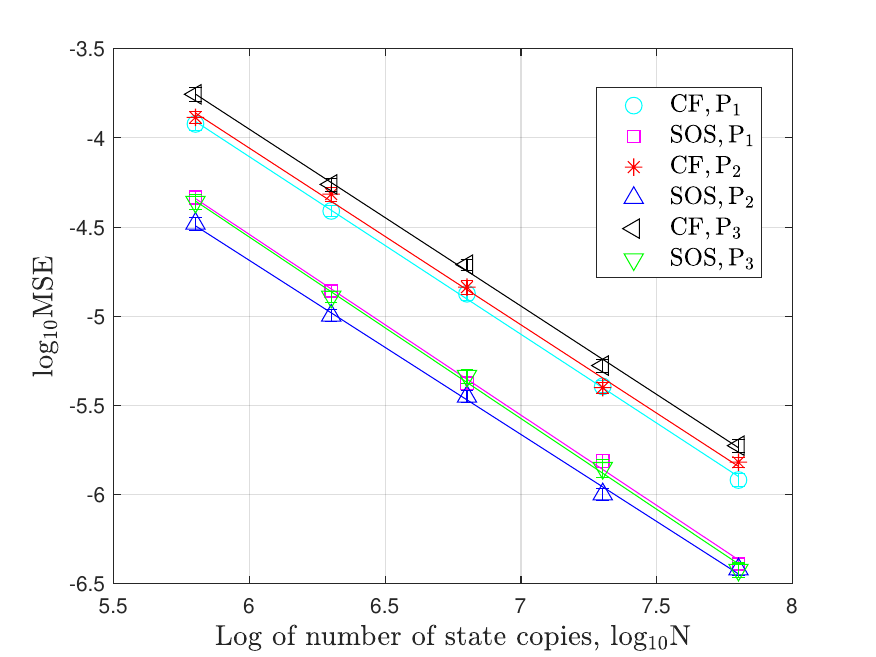}
	\centering{\caption{Log-log plot of MSE versus the total number of state copies $N$  for  I-QDT using the closed-form (CF) solution and SOS optimization with $M=20$ random probe states. }\label{f4}}
\end{figure}
\begin{figure}[!t]
	\centering
	\includegraphics[width=3.3in]{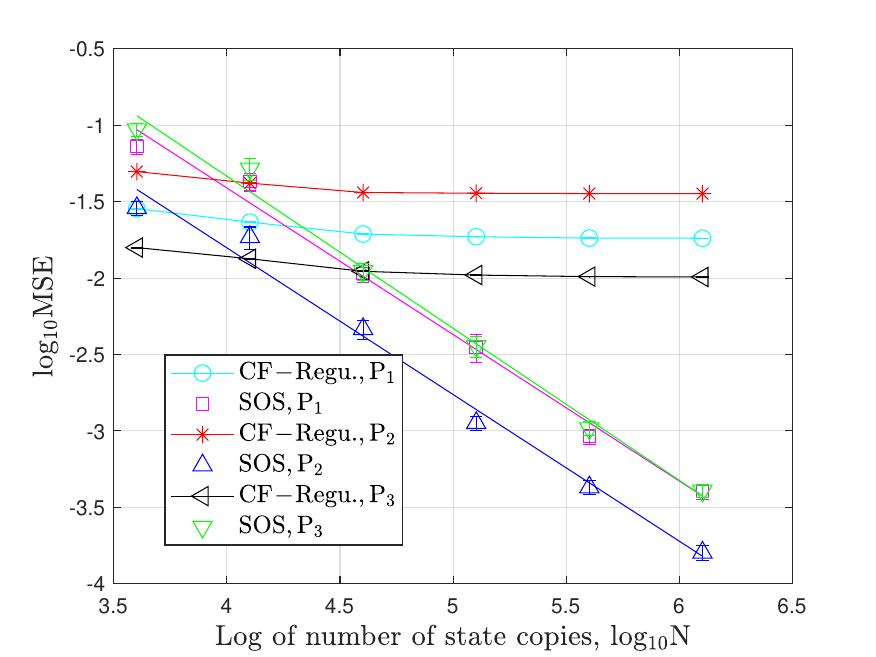}
	\centering{\caption{Log-log plot of MSE versus  the total number of state copies $N$  for   I-QDT using the closed-form solution with regularization (CF-Regu.)  and SOS optimization with $M=4$ random probe states. }\label{f5}}
\end{figure}

Here we focus on I-QDT for collective QDT, where we consider
a three-valued detector $(L=3)$ defined as
\begin{equation}\label{onede}
	\begin{aligned}
		P_{1}&=U_{1}\operatorname{diag}\big(0.4,0.1\big)U_{1}^{\dagger}, \\
		P_{2}&=U_{2}\operatorname{diag}\big(0.5,0.1\big)U_{2}^{\dagger}, \\
		P_{3}&=I-P_{1}-P_{2}\geq 0,
\end{aligned}\end{equation}
where $ U_1 $ and $ U_2 $ are randomly generated unitary matrices and then fixed in the simulation. For this one-qubit example using SOS optimization, the condition $\phi_l \in \mathcal{K}$ is equivalent to $\sum_{i=1}^{4} (\phi_l)_i^2 \leq 1$. Thus,
the SOS optimization problem can be formulated as
\begin{equation}\label{sosqdone}
	\begin{aligned}
		\min&\;(-\gamma)\\
		\text{s.t. } &\sum_{l,k=1}^{4,4}||\hat Y_{lk}-\Phi\left(\phi_l\otimes \phi_k\right)||^2-\gamma \text{ is SOS,}\\
		&\sum_{l=1}^3 \phi_l=[\sqrt 2,0, 0, 0],\\
		& \sum_{i=1}^{4} (\phi_l)_i^2 \leq 1 \text{ for } 1\leq l \leq 3.
	\end{aligned}
\end{equation}

We randomly generate $M=20$ different probe states that are weakly informational-complete and $M=4$ different probe states that are weakly informational-incomplete, and then fix them in the simulation. For a closed-form solution, we add regularization $D=\frac{1000}{N}I$ in the weakly informational-incomplete scenario.

The results are presented in Figs. \ref{f4} and \ref{f5}.
In Fig. \ref{f4}, all MSE scalings follow $O(1/N)$, which is consistent with Theorem \ref{theorem2}.
In the weakly informational-incomplete scenario shown in Fig. \ref{f5}, the MSE of SOS optimization still scales as $O(1/N)$, even when using only four random probe states. For the closed-form solution with regularization, the MSE decreases for $N < 10^5$, and it is smaller than that of the SOS optimization when $N<10^4$. However, as $N$ increases, the MSE remains nearly constant. This behavior arises because, for $N > 10^5$ in the weakly informational-incomplete scenario, the estimates ${\tilde{R}_{ll}}$ in \eqref{dtf} deviate from the true values, which impedes the convergence, as similarly proved in \cite{Xiao2023}.

Table \ref{table1} illustrates the consumption time for both the closed-form solution and SOS optimization corresponding to Figs.~\ref{f4} and~\ref{f5}. The closed-form solution shows significantly lower time consumption compared to SOS optimization. While the closed-form solution is faster, it often sacrifices accuracy. In contrast, SOS optimization offers higher accuracy but requires considerably more computational time. Thus, there is a trade-off between accuracy and consumption time for these two algorithms.
Additionally, the closed-form solution necessitates a large number of distinct probe states to ensure weak informational completeness. However, SOS optimization can achieve high accuracy with a significantly reduced number of probe states.

\begin{table}
	\caption{Time consumption in collective QDT using the closed-form solution in Section \ref{closed} and SOS optimization in  \eqref{sosqdone}.}
	\renewcommand{\arraystretch}{1.4}
	\label{table1} 
	
	\centering
	
	\begin{tabular}{|p{1.9cm}<{\centering}|p{3.3cm}<{\centering}|p{2cm}<{\centering}|}
		\toprule
		The setting&The closed-form solution &SOS\\
		\hline
		$M=20$ & $0.226$ sec &$4527.843 $ sec \\
		\hline
		$M=4$& $0.295$ sec  & $4423.916  $ sec\\
		\bottomrule
	\end{tabular}
	
\end{table}

\begin{figure}
	\centering
	\includegraphics[width=3.3in]{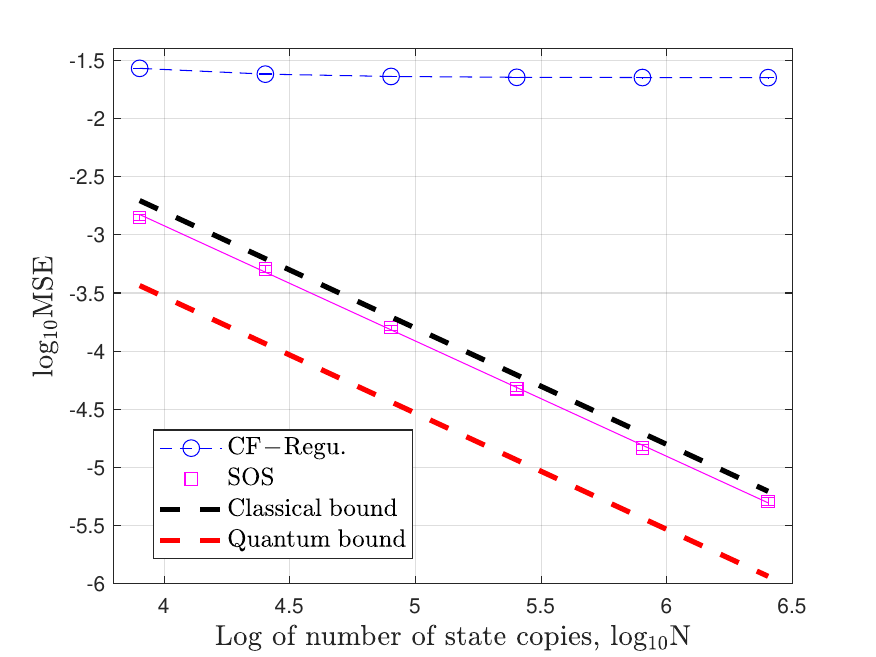}
	\centering{\caption{Log-log plot of MSE versus the total number of state copies $N$, comparing the closed-form algorithm with regularization (CF-Regu.), the SOS optimization method, and the classical and quantum detector bounds from~\cite{dasbound}.}\label{fg2}}
\end{figure}

Moreover, probing with the last eight MUB states from the MUB basis of~\eqref{mubbase}, we compare our results against theoretical precision bounds for QDT proposed in~\cite{dasbound}. These theoretical bounds, stemming from the Cram\'{e}r-Rao theorem, leverage quantum-statistical principles to lower-bound the minimum MSE achievable in unbiased detector estimation. The classical Cram\'{e}r-Rao bound captures the minimum MSE for given probe states, whereas the quantum Cram\'{e}r-Rao bound captures the minimum MSE for the optimal probe states. The result of our comparison is shown in Fig.~\ref{fg2}. Our SOS solution is close to the classical Cram\'{e}r-Rao bound, given by
\[
\log_{10} \operatorname{MSE} = 1.1978 - \log_{10} N \, .
\] 
Interestingly, the SOS solution is observed to be even lower than the classical bound. This effect may be attributed to bias in the estimates using our algorithms, which can lead to a lower MSE. In contrast, there remains a gap between our results and the quantum Cram\'{e}r-Rao bound, given by
\[
\log_{10} \operatorname{MSE} = 0.4669 - \log_{10} N .
\] 
However, it remains an open question as to whether the quantum bound, which was proven tight for phase-insensitive measurements, is attainable in this case~\cite{dasbound}.


\subsubsection{Two-copy collective bit–phase flip process tomography}
For collective QPT,
we consider  the bit–phase flip process \cite{qci} which is TP. The corresponding Kraus operators are given by
\begin{equation}
A_1=\sqrt{p}\left[\begin{array}{cc}{1}&{0}\\{0}&{1}\end{array}\right], \quad A_2=\sqrt{1-p}\left[\begin{array}{cc}{0}&{-\mathrm{i}}\\{\mathrm{i}}&{0}\end{array}\right],
\end{equation}
where $1-p$ denotes the probability that the qubit undergoes a bit–phase flip error~\cite{qci}.

\begin{figure}
	\centering
	\includegraphics[width=3.3in]{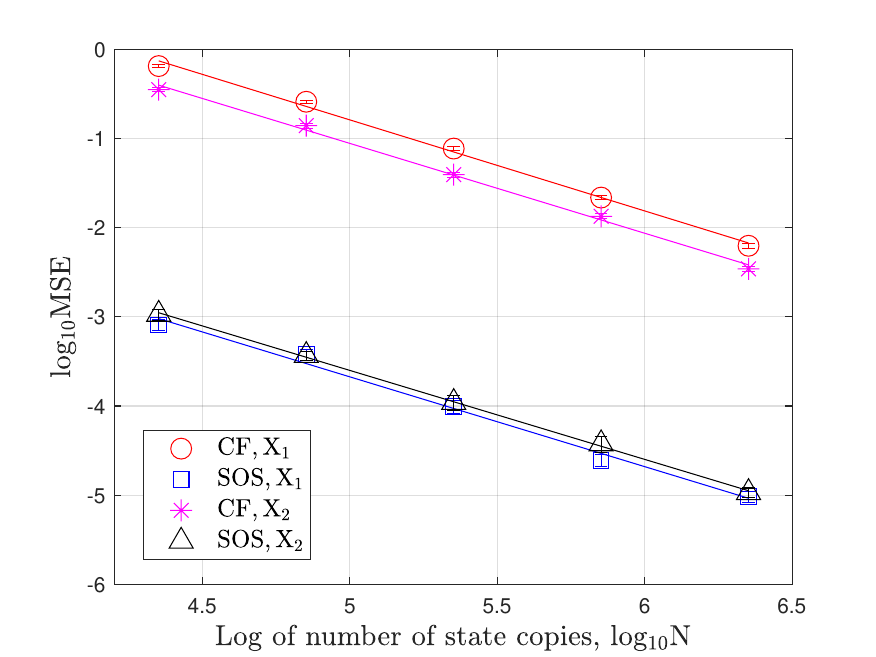}
	\centering{\caption{Log-log plot of MSE versus  the total number of state copies $N$  for   D-QPT using the closed-form (CF) solution and SOS optimization, with $M=16$ random probe states and MUB measurements. }\label{f6}}
\end{figure}

For D-QPT, we set $p_1=0.8$ for the process $\mathcal{E}_1$ and $p_2=0.7$ for the process $\mathcal{E}_2$, and construct the corresponding process matrices $X_1$, $X_2$.
We input $16$ randomly generated and then fixed different quantum states and the measurement is MUB in \eqref{mubbase}.
In the closed-form solution, we first reconstruct the full process matrix $\tilde X$ and decouple it to give $\tilde{X}_1$ and $\tilde{X}_2$, from which we 
 determine the optimal flipping probabilities $\hat p_1 $ and $\hat p_2$. Subsequently, we reconstruct the estimated process matrices $\hat X_1$ and $\hat X_2$ using $\hat p_1$ and $\hat p_2$.
In contrast, the SOS optimization method directly identifies the probability parameters $p_1$ and $p_2$.
The results, presented in Fig. \ref{f6}, show that the MSE scalings are $O(1/N)$, in agreement with Theorem \ref{theorem3}.
Additionally, the MSE of the SOS optimization is significantly smaller than that of the closed-form solution. This is because the closed-form solution does not fully leverage the structural properties of the process matrix in the optimization.

\begin{figure}
	\centering
	\includegraphics[width=3.3in]{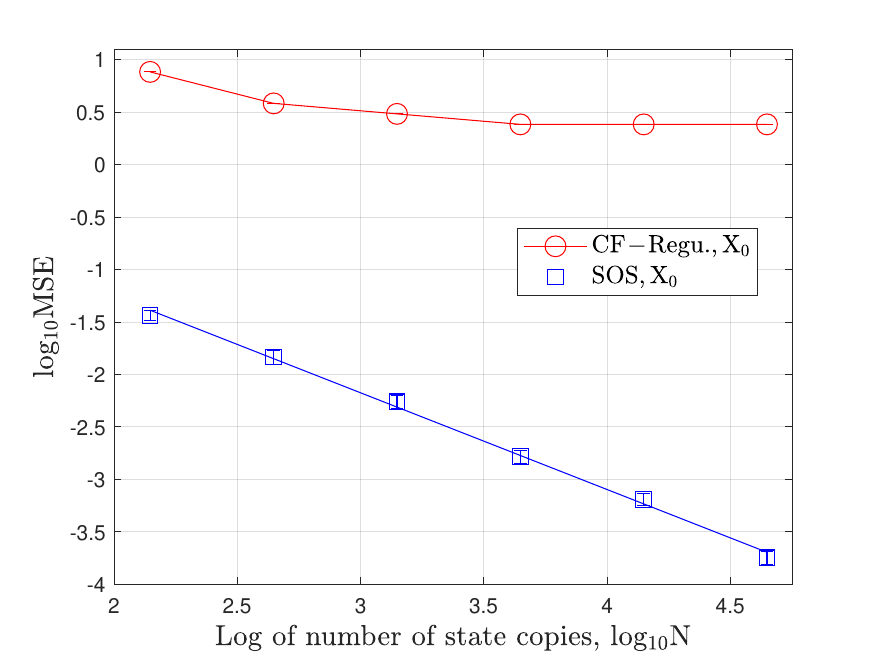}
	\centering{\caption{Log-log plot of MSE versus the total number of state copies $N$  for  I-QPT using the closed-form solution with regularization (CF-Regu.) and SOS optimization, with only $M=1$ random probe state and MUB measurements~\eqref{mubbase}. }\label{f7}}
\end{figure}

For I-QPT, we set $p=0.8$ and construct the corresponding process matrix $X_0$. In this case, we input only one random state. Under the same conditions as D-QPT, the results are presented in Fig. \ref{f7}. The MSE scaling for SOS optimization remains $O(1/N)$, while the MSE of the closed-form solution with regularization does not change when $N\gtrsim10^{3.5}$.  Results indicate that SOS optimization is particularly suitable for problems with structural properties and a smaller number of unknown parameters.


\section{Experimental results}\label{sec7}

Here
we validate the proposed algorithms using experimental data from \cite{Hou2018}, which implemented a two-copy collective measurement in QST via photonic quantum walks. The two copies are the same, which is our I-QST case and was also demonstrated in~\cite{11158864}.
The special two-copy collective measurement was proposed in \cite{PhysRevLett.120.030404,Hou2018} and there are five POVM elements as presented in \eqref{21} and \eqref{22} in Appendix \ref{appb}.

\begin{figure}
	\includegraphics[width=3.3in]{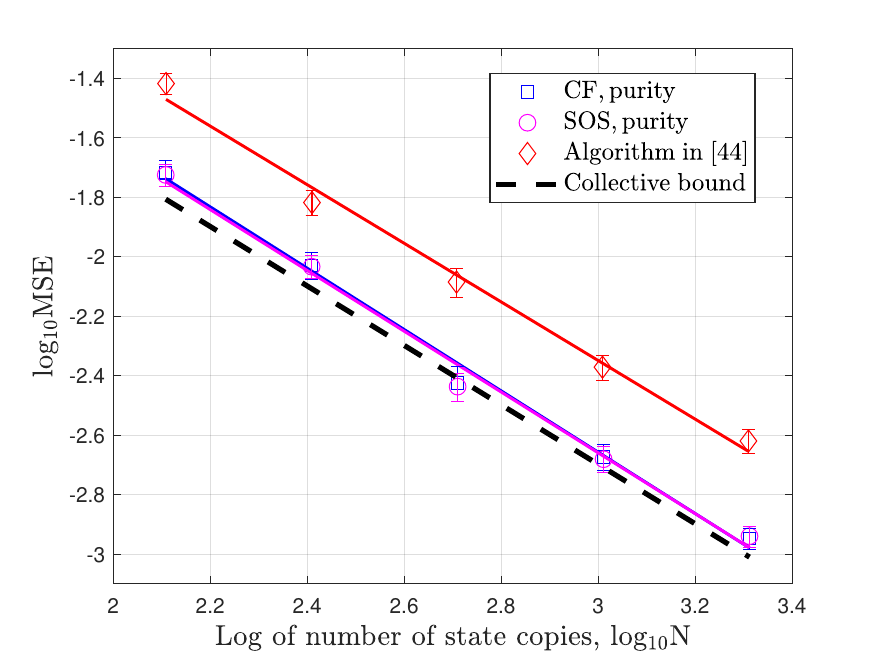}
	{\caption{\!\!\!\!\cite{11158864}~Log-log plot of MSE versus the total number of state copies $N$ based on the experimental data from \cite{Hou2018}. We compare the performance of the modified closed-form (CF) solution, SOS optimization, and the algorithm proposed in \cite{Hou2018}, all implemented with two-copy collective measurements as described in \eqref{21} and \eqref{22}.}\label{ff2}}
\end{figure}

For a qubit state $\rho =\frac{1}{2}I_2 + \theta_{2} \frac{\sigma_x}{\sqrt{2}}+\theta_{3} \frac{\sigma_y}{\sqrt{2}}+\theta_{4} \frac{\sigma_z}{\sqrt{2}}$,  the corresponding measurement result of $P_5^{(2)}$  can be calculated as
\begin{equation}\label{non}
	\begin{aligned}
		p_5=\operatorname{Tr}(P_5^{(2)}\rho^{\otimes 2}) =\frac{1}{4}-\sum_{i=2}^{4} \frac{\theta_{i}^2}{2}.
	\end{aligned}
\end{equation}
The purity of $\rho$ is $\operatorname{Tr}(\rho^2)=\frac{1}{2}+\sum_{i=2}^{4} \theta_i^2$ and thus $1-2	\hat p_5$ can be directly recognized as an estimate of $\operatorname{Tr}(\rho^2)$.
Using collective measurement, the measurement data $1-2	\hat p_5$ provides a direct estimate of purity, while for individual tomography of a qubit, the purity cannot be directly estimated from the measurement data.  In fact, with individual measurements, we can only obtain linear information about the unknown parameters. This is because the measurement data can always be expressed as $\hat{p}=\operatorname{Tr}(P_l\rho)=\sum_{i=1}^{4} \phi_i \theta_{i}$.  In contrast, collective measurements allow us to obtain nonlinear information about the unknown parameters \cite{poly}, as the purity of the qubit illustrated in \eqref{non} \cite{PhysRevA.75.012104}. This analysis highlights the significant role of entanglement in enhancing the efficiency of quantum tomography.
 
Based on this collective measurement, we modify our closed-form solution and SOS optimization.
Using POVM elements $P_l^{(2)}$ for $1\leq l \leq 4$, we have
\begin{equation}
p_l=\operatorname{Tr}  \left( \frac{3}{4}\left(\left|\psi_l\right\rangle\left\langle\psi_l\right|\right)^{\otimes2} \rho^{\otimes 2}\right)=\frac{3}{4} \left(\left\langle\psi_l\right|\rho \left|\psi_l\right\rangle\right)^2.
\end{equation}
Therefore, using measurement data $\{\hat {p}_l\}_{l=1}^{4}$ and LRE \cite{Qi2013}, we obtain a unique estimate, expressed as $\tilde \rho$. Let $$\tilde\rho =\frac{1}{2}I_2 + \tilde\theta_{2} \frac{\sigma_x}{\sqrt{2}}+\tilde\theta_{3} \frac{\sigma_y}{\sqrt{2}}+\tilde\theta_{4} \frac{\sigma_z}{\sqrt{2}}.$$
We aim to determine
 $$\hat\rho =\frac{1}{2}I_2 + \hat\theta_{2} \frac{\sigma_x}{\sqrt{2}}+\hat\theta_{3} \frac{\sigma_y}{\sqrt{2}}+\hat\theta_{4} \frac{\sigma_z}{\sqrt{2}}$$ by solving the optimization problem:
\begin{equation}
	\begin{aligned}
		\min_{\hat{\theta}_2, \hat{\theta}_3,\hat{\theta}_4}&\; \|\hat\rho -\tilde\rho\|^2=\sum_{i=2}^{4}\left| \hat\theta_i- \tilde\theta_i\right|^2\\
		\text{s.t. } & \sum_{i=2}^{4} \hat{\theta}_i^2 = \frac{1}{2}-2\hat p_5.
	\end{aligned}
\end{equation}
Using the Lagrange multiplier method, the optimal solution is
\begin{equation}
   \hat\theta_{i}= \tilde\theta_{i} \sqrt{\frac{\frac{1}{2}-2\hat p_5}{\sum_{i=2}^4 \tilde \theta_i^2}}, \; 2\leq i \leq 4.
\end{equation}
Since $\operatorname{Tr}(\hat{\rho}^2) = 1 - 2\hat{p}_5 \leq 1$, $\hat{\rho}$ satisfies the positive semidefiniteness constraint, ensuring that the final estimate $\hat{\rho}$ is physical.

We then show that $\mathbb{E}\|\hat \rho -\rho\|^2=O(1/N)$.
Using LRE, we have 
\begin{equation}\label{t1}
   \mathbb{E}\|\tilde \rho -\rho\|^2=O\left(\frac{1}{N}\right), 
\end{equation}
which implies
\begin{equation}\label{pp1}
\mathbb{E}|\operatorname{Tr}(\tilde \rho^2)-\operatorname{Tr}(\rho^2)|=O\left(\frac{1}{N}\right).
\end{equation}
Since $\mathbb{E}|1-2\hat{p}_5-\operatorname{Tr}(\rho^2)|^2=O(1/N)$, combining this with \eqref{pp1}, we have
\begin{equation}
\mathbb{E}\left|1-2\hat{p}_5-\operatorname{Tr}(\tilde \rho^2)\right|^2=\mathbb{E}\Big|\frac{1}{2}-2\hat{p}_5-\sum_{i=1}^{3} \tilde \theta_i^2\Big|^2=O\left(\frac{1}{N}\right).
\end{equation}
Let $\dfrac{\frac{1}{2}-2\hat p_5}{\sum_{i=2}^4 \tilde \theta_i^2}=1+t$ and thus $\mathbb{E}|t|^2=O(1/N)$. We then have
\begin{equation}
\begin{aligned}
    &\mathbb{E}|  \hat\theta_{i}-\tilde\theta_{i}|^2= \mathbb{E}|\sqrt{1+t}-1|^2|\tilde\theta_{i}|^2\\
    \sim& \mathbb{E}\frac{|t|^2}{4}|\tilde\theta_{i}|^2 =O\left(\frac{1}{N}\right), 2\leq i \leq 4,  
\end{aligned}
\end{equation}
and thus
\begin{equation}
   \mathbb{E}\|\hat\rho-\tilde \rho\|^2=\mathbb{E}\sum_{i=2}^4|  \hat\theta_{i}-\tilde\theta_{i}|^2=O\left(\frac{1}{N}\right). 
\end{equation}
Finally, combining this with \eqref{t1}, we obtain
\begin{equation}
   \mathbb{E}\|\hat\rho- \rho\|^2=O\left(\frac{1}{N}\right). 
\end{equation}

Using purity information,
the SOS optimization problem can be modified as:
\begin{equation}\label{sosqsone2}
	\begin{aligned}
		\min&\;(-\gamma)\\
		\text{s.t. } &||\hat Y-\Phi\left(\theta\otimes \theta\right)||^2-\gamma \text{ is SOS,}\\
		& \theta_1=\frac{1}{\sqrt{2}},\; \sum_{i=2}^{4} \theta_i^2 = \frac{1}{2}-2\hat{p}_5.
	\end{aligned}
\end{equation}

In the experiment described in \cite{Hou2018}, the unknown quantum state is  
\begin{equation}
    \rho = \frac{I_2}{2} + \frac{1}{\sqrt{2}}\frac{\sigma_x}{2} + \frac{1}{\sqrt{2}}\frac{\sigma_z}{2}.
\end{equation}
Collective measurements as defined in \eqref{21} and \eqref{22} were performed on two-copy states, and the experiment was repeated $100$ times for each given number of copies. The total numbers of copies considered are $N=128, 256, 512, 1024,$ and $2048$. When collective measurements on two identical qubits are allowed, the achievable precision is constrained by a collective bound. The lower bound for the MSE was given in \cite{zhu,Hou2018} as  
\begin{equation}\label{collb}
	\left\{
	\begin{array}{lll}
		\frac{\big(2+\sqrt{1-s^2}\big)^2}{3N}, & \quad 0 \leq s \leq \frac{3+4\sqrt{3}}{13}, \\[10pt]
		\frac{s(1+s)(3-s)}{(3s-1)N}, & \quad \frac{3+4\sqrt{3}}{13} \leq s \leq 1,
	\end{array}
	\right.
\end{equation}
where $s = \sqrt{2\sum_{i=2}^{4}\theta_i^2}$.  

Ref.~\cite{Hou2018} proposed a modified accelerated projected-gradient algorithm for state reconstruction. In contrast, we employed our modified closed-form solution together with SOS optimization techniques. The results, shown in Fig.~\ref{ff2}, demonstrate that both methods outperform the algorithm of~\cite{Hou2018}, with MSE scalings approximately $O(1/N)$. Furthermore, our approaches closely follow the collective bound \eqref{collb}, underscoring their effectiveness for quantum state reconstruction.

\section{Conclusion} \label{sec8}
In this paper, we extended collective quantum state tomography to a generalized collective framework encompassing quantum state, detector, and process tomography. These tasks were formulated as optimization problems, and we developed a closed-form algorithm with explicit computational complexity and MSE scalings. In addition, we reformulated the problems as SOS optimization with semi-algebraic constraints, which, except for the D-QST case, achieve higher accuracy at the cost of increased computation time. We also investigated several illustrative examples, including pure quantum states, projective measurements, and unitary processes.  

The proposed methods were validated through numerical simulations and further tested using experimental data, where collective measurements provided additional purity information about the state. Compared to previous algorithms, our methods can achieve lower MSEs and approach the collective MSE bound. Future work will study the adaptivity of the collective tomography algorithms and extend their applicability to larger-scale and more complex quantum systems.


\appendices
\section{Several lemmas}\label{appendixa}
\begin{lemma}\label{pro1}
(Theorem 10 on Page 55 of \cite{magnus2019matrix})
Let $A$ be an $m \times n$ matrix and $B$ a $p \times q$ matrix, then
\begin{equation}\label{lemma4}
	\operatorname{vec}(A \otimes B)=\left(I_n \otimes K_{q m} \otimes I_p\right)(\operatorname{vec} (A) \otimes \operatorname{vec} (B)),
\end{equation}
where $  K_{q m} $ is a commutation matrix such that $ K_{q m} \operatorname{vec}(O)=\operatorname{vec}(O^{T}) $ and $ O $ is a $ q \times m $ matrix.
\end{lemma}

\begin{lemma}(\cite{bhatia2007perturbation}, Theorem 8.1)\label{lemma1}
	Let $X$, $Y$ be Hermitian matrices with eigenvalues $\lambda_1(X)\geq\cdots\geq\lambda_n(X)$ and $\lambda_1(Y)\geq\cdots\geq\lambda_n(Y)$, respectively. Then
	\begin{equation}\label{weyl}
		\max_j|\lambda_j(X)-\lambda_j(Y)|\leq||X-Y||.
	\end{equation}
\end{lemma}
	\begin{lemma}\cite{dbound}\label{lemma2}
	Let $\mathbb{H}_A$ and $\mathbb{H}_B$ be finite-dimensional Hilbert spaces of dimensions $d_{A}$ and $d_{B}$, respectively, and let $X \in \mathbb{H}_A \otimes \mathbb{H}_B$. Then for any unitarily invariant norm that is multiplicative over tensor products, the partial trace satisfies the norm inequality
\begin{equation}
\left\|\operatorname{Tr}_{A}(X) \right\| \leq \frac{d_{A}}{\left\|I_{A}\right\|}\|X\|,
\end{equation}
where $I_{A}$ is the identity operator.
\end{lemma}

\begin{lemma}[Theorem 2 in \cite{45105560}]\label{lemma5}
For states $\rho$ and $\sigma$ with $\Delta = \rho - \sigma$, 
$T = \|\Delta\|$ and $\beta = \lambda_{\min}(\sigma)$, we have
\begin{equation}
   D(\rho\|\sigma) \leq \frac{T^{2}}{\beta}.
\end{equation}
where  $D(\rho\|\sigma)=\operatorname{Tr}\left( \rho ( \log \rho - \log \sigma ) \right))$ is the quantum relative entropy.
\end{lemma}




\section{MUB and collective measurements}\label{appb}
For $ d=4 $, five MUB measurement sets \cite{mubreview}  are
\begin{small}
	\begin{equation}
		\left\{\!|\psi_{n}^{(\text{MUB})}\rangle\!\right\}\!=\!\left\{\!\left\{\left|\psi_{n}^{A}\right\rangle\!\right\}\!,\!\left\{\left|\psi_{n}^{B}\right\rangle\!\right\}\!,\!\left\{\left|\psi_{n}^{C}\right\rangle\!\right\}\!,\!\left\{\left|\psi_{n}^{D}\right\rangle\!\right\}\!,\!\left\{\left|\psi_{n}^{E}\right\rangle\!\right\}\!\right\},
	\end{equation}
\end{small}
and
\begin{align*}
	&\left\{\left|\psi_{n}^{A}\right\rangle\right\}=\{|00\rangle,|01\rangle,|10\rangle,|11\rangle\},\\
	&\left\{\left|\psi_{n}^{B}\right\rangle\right\}=\{|R\pm\rangle,|L\pm\rangle\},\\
	&\left\{\left|\psi_{n}^{C}\right\rangle\right\}=\{|\pm R\rangle,|\pm L\rangle\},\\
	&\left\{\left|\psi_{n}^{D}\right\rangle\right\}=\left\{\frac{1}{\sqrt{2}}(|R 0\rangle \pm \mathrm i|L 1\rangle), \frac{1}{\sqrt{2}}(|R 1\rangle \pm \mathrm i|L 0\rangle)\right\},\\
	&\left\{\left|\psi_{n}^{E}\right\rangle\right\}=\left\{\frac{1}{\sqrt{2}}(|R R\rangle \pm \mathrm i|L L\rangle), \frac{1}{\sqrt{2}}(|R L\rangle \pm \mathrm i|L R\rangle)\right\}, \numberthis \label{mubbase}
\end{align*}
where $|\pm\rangle=(|0\rangle \pm|1\rangle) / \sqrt{2}$, $|R\rangle=(|0\rangle-{\mathrm i}|1\rangle) / \sqrt{2}$, and
$|L\rangle=(|0\rangle+{\mathrm i}|1\rangle) / \sqrt{2}$ in the natural basis. In this paper, we call $ \rho_n=|\psi_{n}^{(\text{MUB})}\rangle\langle\psi_{n}^{(\text{MUB})}| $  a MUB state and $ P_n=|\psi_{n}^{(\text{MUB})}\rangle\langle\psi_{n}^{(\text{MUB})}| $  a MUB measurement operator.

For one-qubit system, the SIC-POVM \cite{sic} are 
\begin{equation}\label{sic}
	\begin{aligned}
	\left|\psi_{1}\right\rangle&=|0\rangle,\\
	\left|\psi_{2}\right\rangle&=\frac{1}{\sqrt{3}}\big(|0\rangle+\sqrt{2}|1\rangle\big),\\
	\left|\psi_{3}\right\rangle&=\frac{1}{\sqrt{3}}\big(|0\rangle+{e}^{\frac{2\pi\mathrm{i}}{3}}\sqrt{2}|1\rangle\big),\\
	\left|\psi_{4}\right\rangle&=\frac{1}{\sqrt{3}}\big(|0\rangle+{e}^{-\frac{2\pi\mathrm{i}}{3}}\sqrt{2}|1\rangle\big).
	\end{aligned}
	\end{equation}
where the corresponding POVMs are $P_l^{(1)}=\left|\psi_{l}\right\rangle \langle\psi_{l}|, 1\leq l\leq 4  $. Geometrically, the Bloch vectors
of  $\left|\psi_{l}\right\rangle$
form a regular tetrahedron inside the
Bloch sphere, and $\left|\left\langle\psi_{l}|\psi_{k}\right\rangle\right|^{2}=(2\delta_{lk}+1)/3$.

For two-copy collective QST, a special two-copy collective measurement was proposed in \cite{PhysRevLett.120.030404,Hou2018} and  there are five POVM elements
\begin{equation}\label{21}
	P_l^{(2)}=\frac{3}{4}\left(\left|\psi_l\right\rangle\left\langle\psi_l\right|\right)^{\otimes2}, 1\leq l \leq 4
\end{equation}
and 
\begin{equation}\label{22}
	P_5^{(2)}= |\Psi_{-}\rangle\langle\Psi_{-}|, \; |\Psi_{-}\rangle=\frac1{\sqrt{2}}(|01\rangle-|10\rangle),
\end{equation}
where $\sum_{l=1}^{5} P_l =I$. Therefore, the POVM elements $\{P_l^{(2)}\}_{l=1}^{4}$ are all product measurements, while only $P_5^{(2)}$ is the singlet, which is maximally
entangled. 

For three-copy collective QST,  a special three-copy collective measurement was proposed in \cite{zhou2023experimental} and  there are seven POVM elements
\begin{equation}\label{31}
	P_l^{(3)}=\frac{2}{3}|\varphi_l\rangle\langle\varphi_l|^{\otimes3},l=1,\cdots,6,\; P_7^{(3)}={I}-\sum_{l=1}^6 P_l^{(3)},
\end{equation}
where 
\begin{equation}
	\begin{aligned}
		|\varphi_{1}\rangle&=|0\rangle,\;|\varphi_{2}\rangle  =|1\rangle,  \\
		|\varphi_{3}\rangle&=\frac{1}{\sqrt{2}}(|0\rangle+|1\rangle),\;|\varphi_{4}\rangle  =\frac1{\sqrt{2}}(|0\rangle-|1\rangle),  \\
		|\varphi_{5}\rangle&=\frac{1}{\sqrt{2}}(|0\rangle+\mathrm{i}|1\rangle),\;|\varphi_{6}\rangle=\frac{1}{\sqrt{2}}(|0\rangle-\mathrm{i}|1\rangle).	
	\end{aligned}
\end{equation}
These six vectors  form a regular octahedron
when represented on the Bloch sphere.  Therefore, the POVM elements $\{P_l^{(3)}\}_{l=1}^{6}$ are all product measurements, while only $P_7^{(3)}$ is the entangled measurement.

\bibliographystyle{ieeetr}         
\bibliography{collective} 

\begin{IEEEbiographynophoto}{Shuixin Xiao} (Member,~IEEE) received a B.E. degree in automation from Huazhong University of Science and Technology, Wuhan, China, in 2018, and a Ph.D. degree in control theory and engineering from Shanghai Jiao Tong University, Shanghai, China, in 2023. He also received a Ph.D. degree in electrical engineering from the University of New South Wales, Canberra, Australia, in 2023. He was Post-Doctoral Fellow with the School of Engineering and Technology, University of New South Wales, Canberra, Australia and School of Engineering, Australian National University between 2023 and 2025.
He is now a Research Fellow with the Department of Electrical and Electronic Engineering, University of Melbourne, Australia. His current research interests include quantum system identification, quantum control and quantum information theory. 
\end{IEEEbiographynophoto}
\begin{IEEEbiographynophoto}{Yuanlong Wang} (Senior Member,~IEEE) received a B.E. degree in automation
	from Northeastern University, Shenyang, China, in
	2011, an M.E. degree in control science and engineering
	from Zhejiang University, Hangzhou, China, in 2015,
	and a Ph.D. degree in engineering from the University
	of New South Wales, Canberra, ACT, Australia, in 2019.
	He was a Post-Doctoral Fellow with the Centre for
	Quantum Dynamics, Griffith University, Brisbane, QLD,
	Australia, from 2019 to 2022. He is now an Associate
	Professor in the Academy of Mathematics and Systems
	Science, Chinese Academy of Sciences, Beijing, China.
	His research interests include quantum control theory and quantum parameter
	estimation theory.
\end{IEEEbiographynophoto}

\begin{IEEEbiographynophoto}{Zhibo Hou} received a B.S. degree in applied physics
from Beijing University of Posts and Telecommunications, Beijing, China, in 2011, and a Ph.D. degree in Optics
from the University of Science and Technology of China
(USTC), Hefei, China, in 2016. From 20016 to 2018, he was
a post-doctoral fellow with the School of Physics, USTC.
Currently, he is a professor at the University
of Science and Technology of China, Hefei, China.
His research interests include quantum control,
quantum metrology, quantum tomography, and experimental quantum optics.

\end{IEEEbiographynophoto}

\begin{IEEEbiographynophoto}{Aritra Das} received a B.S. degree
in Physics from the Indian Institute of Technology, Kanpur, India
in 2020. He started his Ph.D. degree in 2021
at the Department of Quantum Science and Technology
in the Australian National University
and is currently
pursuing it. His research interests include
quantum information and quantum metrology,
especially quantum estimation, quantum sensing
and resource theories.
\end{IEEEbiographynophoto}

\begin{IEEEbiographynophoto}{Ian R. Petersen} (Life Fellow,~IEEE)  was born in Victoria, Australia. He received a Ph.D. in Electrical Engineering in 1984 from the
	University of Rochester, USA. From 1983 to 1985 he was a Postdoctoral Fellow at the Australian National University.
	In 1985 he joined the University of New South Wales, Canberra, Australia. He moved to The Australian National
	University in 2017 where he is currently a Professor in the  School of Engineering. He was the Australian
	Research Council Executive Director for Mathematics, Information and Communications in 2002 and 2003. 
	He was Acting Deputy Vice-Chancellor Research for the University of New South Wales in 2004 and 2005. 
	He held an Australian Research Council Professorial Fellowship from 2005 to 2007, an Australian Research Council Federation Fellowship 
	from 2007 to 2012, and an Australian Research Council Laureate Fellowship from 2012 to 2017.
	
	He served as an Associate Editor for the IEEE Transactions on Automatic
	Control, Systems and Control Letters, Automatica, and SIAM Journal on Control and
	Optimization. He was an Editor for Automatica in the area of optimization
	in systems and control. He is a fellow of IFAC, the IEEE and the Australian Academy
	of Science.
	
	His main research interests are in robust control theory, quantum control theory
	and stochastic control theory. Ian Petersen was elected IFAC Council Member for
	the 2014–2017 Triennium. He was also elected to be a member of the IEEE Control
	Systems Society Board of Governors for the periods 2011–2013 and 2015–2017. He was a Vice President for Technical Activity for the IEEE Control
	Systems Society for 2023--2024. He
	was Vice President for Technical Activity for the Asian Control Association and was
	General Chair of the 2012 Australia Control Conference. He was General Chair of
	the 2015 IEEE Multi-Conference on Systems and Control.
\end{IEEEbiographynophoto}

\begin{IEEEbiographynophoto}{Farhad Farokhi} (Senior Member,~IEEE) received the Ph.D. degree in automatic control from the KTH Royal Institute of Technology, Stockholm, Sweden, in 2014. He joined the University of Melbourne, Parkville, VIC, Australia, where he is currently a Senior Lecturer (equivalent to Assistant/Associate Professor in North America). From 2018 to 2020, he was also a Research Scientist with the CSIRO's Data61, Canberra ACT, Australia. He has been involved in multiple projects on data privacy and cyber-security funded by the Australian Research Council, the Defence Science and Technology Group, the Department of the Prime Minister and Cabinet, the Department of Environment and Energy, and the CSIRO. Dr. Farokhi was the recipient of the VESKI Victoria Fellowship from the Victoria State Government, Australia, and the McKenzie Fellowship, the 2015 Early Career Researcher Award, and MSE Excellence Award for Early Career Research from The University of Melbourne. He is the Associate Editor for IET Smart Grid, Results in Control and Optimization, and Conference Editorial Board of IEEE Control System Society.
\end{IEEEbiographynophoto}

\begin{IEEEbiographynophoto}{Guo-Yong Xiang} received a B.E. Degree in Physics Education from Anhui Normal University in 2000 and a Ph.D.
degree in Optics from University of Science and Technology
of China (USTC) in 2005, respectively. Currently, he is a
professor at USTC, Hefei, China. He was a research fellow
at Griffith University, Brisbane, Australia from 2007 to
2010.
His research interests include quantum optics and
quantum information, especially quantum precision
measurement, quantum state estimation and quantum
feedback control.
\end{IEEEbiographynophoto}

\begin{IEEEbiographynophoto}{Jie Zhao} is a senior Research Fellow in the Department of Quantum Science and Technology at the Australian National University. She received her PhD from the Australian National University, where her research focused on enhancing quantum information through post-selection techniques. Following her PhD, she joined Xanadu Quantum Technologies as a permanent researcher, working on the development of large-scale higher-order quantum entanglement using time-domain multiplexing technologies—a circuit that provided the platform for the demonstration of quantum supremacy. She then continued her research at the Joint Quantum Institute in a Nobel Prize winning group, where she investigated quantum information processing and laser phase locking based on four-wave mixing in rubidium vapours. Her research interests include quantum optics, quantum metrology, quantum communications, and optical quantum computing. 
\end{IEEEbiographynophoto}

\begin{IEEEbiographynophoto}{Daoyi Dong} (Fellow,~IEEE) is currently a Professor and an ARC Future Fellow at the University of Technology Sydney (UTS).
	He received a B.E. degree in automatic control
	and a Ph.D. degree in engineering from the University
	of Science and Technology of China, Hefei, China, in
	2001 and 2006, respectively. He had visiting positions at
	Princeton University, NJ, USA, University of Melbourne, Australia, RIKEN, Wako-Shi, Japan,
	University of Duisburg–Essen, Germany and The University
	of Hong Kong, Hong Kong.
	
	His research interests include quantum control, quantum estimation and machine learning. He was awarded an ACA Temasek Young Educator Award by The Asian Control
	Association and is a recipient of a Future Fellowship, an International Collaboration Award and an
	Australian Post-Doctoral Fellowship from the Australian Research Council, and a
	Humboldt Research Fellowship from the Alexander von Humboldt Foundation of
	Germany. He is the Founding Chair of Technical Committee on Quantum Computing, Systems and Control, IEEE Control Systems Society. He is a member of Board of Governors, IEEE Control Systems Society and a Vice President of IEEE Systems, Man and Cybernetics Society. He served as an Associate Editor of IEEE Transactions on Neural
	Networks and Learning Systems and a Technical Editor of IEEE/ASME
	Transactions on Mechatronics. He is currently an Associate
	Editor of Automatica and IEEE Transactions on Cybernetics.
\end{IEEEbiographynophoto}

\end{document}